\documentclass[fleqn,usenatbib]{mnras}

\usepackage{newtxtext,newtxmath}

\usepackage[T1]{fontenc}
\usepackage{ae,aecompl,times}

\usepackage{graphbox,graphicx}	
\usepackage{amsmath}	
\usepackage{multirow}
\usepackage[normalem]{ulem}
\usepackage{xcolor}
\usepackage{bm}
\usepackage{lipsum}

\newcommand{\Gpch}{\,h^{-1}{\rm Gpc}}
\newcommand{\Mpch}{\,h^{-1}{\rm Mpc}}

\newcommand{\kpch}{\,h^{-1}{\rm kpc}}
\newcommand{\Mpchc}{\,h^{-3}{\rm Mpc}^3}

\newcommand{\hMpc}{\,h\,{\rm Mpc}^{-1}}
\newcommand{\hMpcc}{\,h^3{\rm Mpc}^{-3}}
\newcommand{\Msh}{\,h^{-1}{\rm M}_\odot}
\newcommand{\Msyr}{\,{\rm M}_\odot {\rm yr}^{-1}}

\renewcommand{\vec}[1]{ {\bf #1} }

\title[MTNG with massive neutrinos]{The MillenniumTNG Project: Impact of massive neutrinos on the cosmic large-scale structure and the distribution of galaxies}
\author[C.~Hern\'andez-Aguayo et al.]{%
\parbox{0.99\textwidth}
{%
C\'esar Hern\'andez-Aguayo$^{1,2}$\thanks{E-mail: 
cesarhdz@MPA-Garching.MPG.DE (CH-A)},
Volker Springel$^{1}$\thanks{E-mail: 
vspringel@MPA-Garching.MPG.DE (VS)},
Sownak Bose$^{3}$,
Carlos Frenk$^{3}$,
Adrian Jenkins$^{3}$,\\
Monica Barrera$^{1}$,
Fulvio Ferlito$^{1}$,
R\"udiger Pakmor$^{1}$,
Simon D. M. White$^{1}$,
Lars Hernquist$^{4}$,\\ 
Ana Maria Delgado$^{4}$, 
Rahul Kannan$^{5}$,
and
Boryana Hadzhiyska$^{6,7}$
}
\\
\\%
$^{1}$Max-Planck-Institut f\"ur Astrophysik, Karl-Schwarzschild-Str. 1, D-85748, Garching, Germany\\%
$^{2}$Excellence Cluster ORIGINS, Boltzmannstrasse 2, D-85748 Garching, Germany\\%
$^{3}$Institute for Computational Cosmology, Department of Physics, Durham University, South Road, Durham, DH1 3LE, UK\\%
$^{4}$Harvard-Smithsonian Center for Astrophysics, 60 Garden St, Cambridge, MA 02138, USA\\%
$^{5}$Department of Physics and Astronomy, York University, 4700 Keele Street, Toronto, ON M3J 1P3, Canada\\%
$^{6}$Miller Institute for Basic Research in Science, University of California, Berkeley, CA, 94720, USA\\%
$^{7}$Physics Division, Lawrence Berkeley National Laboratory, Berkeley, CA 94720, USA%
}

\date{Accepted XXX. Received YYY; in original form ZZZ}

\pubyear{2024}

\begin{document}
\label{firstpage}
\pagerange{\pageref{firstpage}--\pageref{lastpage}}
\maketitle

\begin{abstract}
We discuss the cold dark matter plus massive neutrinos simulations of the MillenniumTNG (MTNG) project, which aim to improve understanding of how well ongoing and future large-scale galaxy surveys will measure neutrino masses. Our largest simulations, $3000\,{\rm Mpc}$ on a side, use $10240^3$ particles of mass $m_{p} = 6.66\times 10^{8}\,h^{-1}{\rm M}_\odot$ to represent cold dark matter, and $2560^3$ to represent a population of neutrinos with summed mass $M_\nu = 100$ meV. Smaller volume runs with $\sim 630\,{\rm Mpc}$ also include cases with $M_\nu = 0\,\textrm{and}\, 300\,{\rm meV}$. All simulations are carried out twice using the paired-and-fixed technique for cosmic variance reduction.  We evolve the neutrino component using the particle-based $\delta f$ importance sampling method, which greatly reduces shot noise in the neutrino density field. In addition, we modify the {\small GADGET-4} code to account both for the influence of relativistic and mildly relativistic components on the expansion rate and for non-Newtonian effects on the largest represented simulation scales. This allows us to quantify accurately the impact of neutrinos on basic statistical measures of nonlinear structure formation, such as the matter power spectrum and the halo mass function.  We use semi-analytic models of galaxy formation to predict the galaxy population and its clustering properties as a function of summed neutrino mass, finding significant ($\sim 10\%$) impacts on the cosmic star formation rate history, the galaxy mass function, and the clustering strength. This offers the prospect of identifying combinations of summary statistics that are optimally sensitive to the neutrino mass.
\end{abstract}




\section{Introduction}
\label{sec:intro}
The advent of ever larger and more precise astronomical data sets over the past decades has led to tremendous progress in our understanding of cosmology, culminating in the establishment of a standard model, $\Lambda$CDM, whose basic premises are now subject to precision tests by ongoing and planned experiments \citep{Laureijs:2011gra,LSST:2009,PFS2014,eBOSS:2020yzd,DESI:2023ytc, Mellier2024}. In fact, the forthcoming new data will be so precise that they will not only be able to detect potential deviations from a cosmological constant, but also be sensitive to the impact of baryonic physics (for example from AGN feedback) on the matter distribution. While this can be viewed as an unwelcome complication, it of course also offers a chance for additional tests of our astrophysical understanding of the relevant aspects of galaxy formation.

In a similar vein, cosmological observations are becoming precise enough to detect deviations from the properties ascribed to purely cold dark matter. For example, dark matter may be composed of such a light particle that its thermal motions are non-negligible, causing smaller structures to be wiped out, and such warm dark matter (WDM) remains a much studied alternative to standard cold dark matter (CDM) scenarios \citep[e.g.][]{Bode2001, Viel2013, Lovell2014}. The most extreme version of such a model, where the particles are all truly `hot', i.e.~with large, nearly relativistic thermal velocities at high redshift, is cleared ruled out by observations because the associated top-down scenario for structure formation is inconsistent with data on multiple fronts. This is also why normal neutrinos as the dominant dark matter component were ruled out early on \citep{White1983}.

However, at the same time we know that a significant neutrino background is left behind by big bang nucleosynthesis, and the phenomenon of neutrino flavour conversion implies a lower bound for the summed neutrino masses \citep[e.g.][]{Esteban2020}. This means that neutrinos represent a hot dark matter admixture that must be present in the standard model, and which constitutes at most of order a few percent of all the dark matter according to a variety of cosmological constraints \citep{Archidiacon2017,DESI:2024mwx}. These are, in fact, tighter than those obtained so far from the best  experimental upper bounds on the neutrino masses which come from the KATRIN experiment, based on an examination of the  beta decay of tritium \citep{Aker2022,Katrin:2024tvg}. Their low mass limits the influence of neutrinos on structure growth to the level of a small perturbation that until recently could be safely ignored in  studies of galaxy formation. However, recent improvements in the precision of modern cosmological probes of galaxy clustering and weak lensing have started to transform this assessment. In fact, the impact of neutrinos on large-scale structure may soon become measurable with galaxy surveys, raising the prospect of actually measuring the sum of the neutrino masses and so distinguishing between the normal and inverted mass hierarchies, a fundamentally important issue in particle physics.

Computing the impact of neutrinos on structure formation in the non-linear regime faces substantial technical difficulties, because a straightforward representation  of the neutrino density field through particles is severely affected by shot noise even on large scales. This contrasts with the cold dark matter case, where shot noise is far below the intrinsic power on most scales of interest.  However, in recent years a large number of approximate simulation techniques have been developed that address this issue in a variety of ways. The recent comparative study of \citet{Adamek2023} gives a comprehensive overview of these methods, and compares their accuracy, albeit  over a considerably smaller spatial dynamic range than studied here, where we employ the $\delta f$-technique originally proposed by \citet{Elbers2021}. This is one of the most accurate and conceptually elegant numerical techniques proposed so far. It uses a particle-based representation that is capable of tracking non-linear neutrino clustering without spatial resolution limitations while at the same time greatly reducing shot noise compared to  naive particle-based methods \citep[e.g.][]{Viel2010,Emberson2017}.

We have used the $\delta f$-approach in the neutrino branch of our new MillenniumTNG simulation programme. The primary introductory paper of MTNG \citep{Hernandez-Aguayo2023} already briefly described some initial results from our first large neutrino run, but it deferred a more detailed discussion to later work that we now present in this study. Here we additionally include results from a second very large neutrino simulation also carried out following 1.1 trillion particles in a 3000 Mpc box, and we include calculations at the same resolution but in smaller boxes that vary the neutrino mass. Other MTNG simulations include large dark-matter-only simulations in a 740 Mpc box at different mass resolutions, as well as a high-resolution hydrodynamical model \citep{Pakmor2023} with full galaxy formation physics for the same initial conditions, carried out with the IllustrisTNG physics model \citep{Weinberger2017, Pillepich2018} but with a volume nearly 15 times larger than TNG300. Initial MTNG analysis papers have already looked at weak gravitational lensing \citep{Ferlito2023}, intrinsic galaxy shape alignments \citep{Delgado2023}, galaxy clustering \citep{Bose2023}, predictions for the high-$z$ galaxy population \citep{Kannan2023}, inference of cosmological parameters \citep{Contreras2023}, and mock galaxy catalogue construction \citep{Barrera2023, Hadzhiyska2023a, Hadzhiyska2023b}.

We note that since these first MTNG papers a similarly large neutrino simulation, {\small FLAMINGO-10k}, has been added to the {\small FLAMINGO} project \citep{Schaye2023} and been used to study the clustering of galaxies and quasars at redshift $z\sim 6$ \citep{Pizzati2024}, based on an empirical quasar model. The {\small FLAMINGO-10k} simulation has a very similar mass resolution and particle number to our large runs, but the adopted neutrino mass is different, only 60 meV, thus nicely complementing our own large simulations.  Another very large cosmological simulation of similar mass resolution, the New Worlds Simulation of \citet{Heitmann2024}, recently accounted for neutrinos, but only at the linear level without proper treatment of non-linear neutrino clustering. \cite{Yoshikawa:2020ehd,Yoshikawa:2021qbw} carried out large-scale cosmological simulations with massive neutrinos using a hybrid N-body/Vlasov approach solving explicitly the Boltzmann equation. On the other hand, the {\small TianNu} simulation \citep{Emberson2017,Yu:2016yfe,Chen:2023vsv} co-evolve nearly 3 trillion CDM and neutrino particles in a $1.2\,\Gpch$ box, making it one of the largest cosmological N-body simulations with massive neutrinos performed to date. The {\small MassiveNuS} simulations \citep{Liu:2017now} have included the effects of radiation on the background expansion and varying the summed mass of neutrinos between 0 meV and 600 meV. Moreover, the {\small DEMNUni} \citep{Castorina:2015bma} and the {\small DUSTGRAIN}-pathfinder \citep{Giocoli:2018gqh} simulations have included the effects of massive neutrinos with dynamical dark energy and modified gravity, respectively.
Other very large cosmological simulations, such as the  {\small AbacusSummit} simulations \citep{Maksimova:2021ynf}, have not so far included neutrinos explicitly. Also, we note that galaxy catalogues based on following the physics of galaxy formation semi-analytically, as we do here, are not yet available for these other simulations; rather, some of them typically use empirical models to represent the galaxy population, derived through halo occupation distribution modelling or similar techniques. We thus consider it one of the special and distinguishing features of the MTNG simulations that we can augment them with semi-analytic predictions computed with a novel version of the {\small L-GALAXIES} code, including the production of mock catalogues on seamless lightcone outputs \citep{Barrera2023}.

In this work, we will focus on introducing the neutrino methodology of our simulation set, and on highlighting the impact of neutrinos on selected clustering statistics, both for haloes and for mock galaxy catalogues. Addressing the question which combination of observational measures can most powerfully constrain neutrino masses is, however, beyond the scope of this paper.

Our paper is structured as follows. In Section~\ref{sec:neutrino} we discuss the technical aspects of our neutrino simulation methods, and in Section~\ref{sec:mtng} we specify the simulations analysed here. We then present results for matter clustering in Section~\ref{sec:matter} and for halo statistics in Section~\ref{sec:halo}. This is followed by an analysis of galaxy clustering in Section~\ref{sec:gal}. Finally, we present a summary of our findings and our conclusions in Section~\ref{sec:conc}.

\section{Theoretical and numerical aspects}
\label{sec:neutrino}
While the techniques for standard cosmological N-body simulations are well understood and extensively documented in the literature \citep[see][for a review]{Angulo2022}, many different approaches exist for including neutrinos \citep{Adamek2023}. In order to provide an unambiguous specification of our own work, we here summarise the methods we use for including massive neutrinos in the evolution both of the background cosmology and of cosmic structure, as well as for setting up the appropriate initial conditions.

\subsection{Cosmological background evolution}
\label{sec:cosmo_nu}
Here we recall some basic aspects of neutrino cosmology as needed in our numerical implementation.  To run our neutrino simulations, we have included the evolution of radiation (photons and relativistic massless neutrinos) and of a hot dark matter component in the form of massive neutrinos (which may experience a transition from relativistic to non-relativistic velocities) in the calculation of the expansion history in {\small GADGET-4}. Hence, the cosmic evolution of the density parameters of different matter and energy components in a flat Universe is given by
\begin{equation}\label{eq:O_nu}
\Omega_r(z) + \Omega_{\rm{cb}}(z) + \Omega_\nu(z) + \Omega_\Lambda = 1\,,
\end{equation}
where $\Omega_r$ is the density parameter of the radiation components, $\Omega_{\rm{cb}}$ is the contribution of the cold matter components (cold dark matter plus baryons), $\Omega_\nu$ is the density of the non-relativistic massive neutrinos, $\Omega_\Lambda$ is the density parameter of dark energy in the form of the cosmological constant where $\Omega_\Lambda = 1 - [\Omega_r(z) + \Omega_{\rm{cb}}(z) + \Omega_\nu(z)]$. 

\begin{figure}
 \centering
\includegraphics[width=0.5\textwidth]{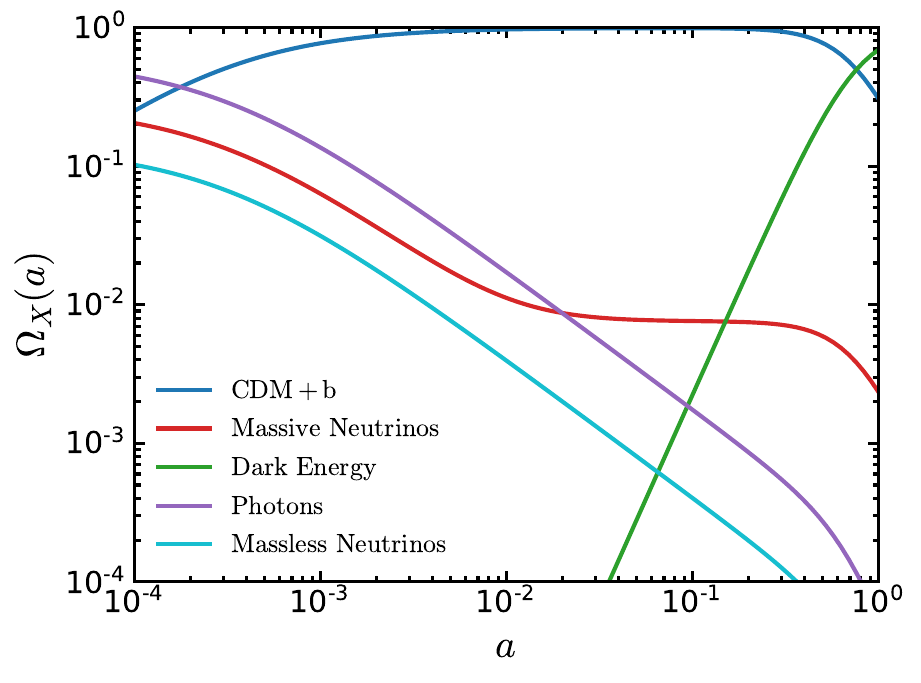}\vspace*{-0.25cm}\\
\caption{Cosmological evolution of the density parameters of different matter and energy components as a function of the scale factor, $a=1/(1+z)$, for the case of two degenerate massive neutrino species with summed rest mass of $M_\nu = 100$ meV and one massless neutrino species. Note how the massive neutrinos (red line) initially drop away with scale factor like a relativistic component, in parallel to photons and massive neutrinos, but then they become non-relativistic and join the evolution of the cold dark matter and baryons below redshifts of $z\sim 100$. The massive neutrinos subsequently make up a constant $\sim 1\%$ fraction of the matter density. At late times, the energy density becomes dominated by the dark energy component.
  \label{fig:Omegas}}
\end{figure}

\begin{figure*}
 \centering
\includegraphics[width=0.47\textwidth]{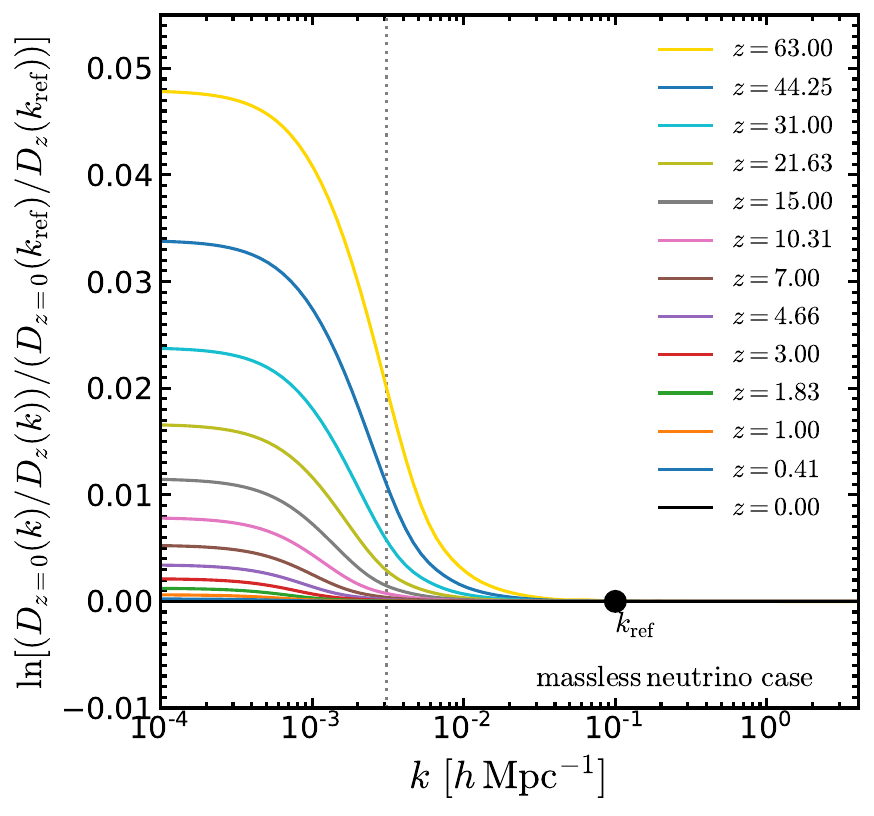}
\includegraphics[width=0.47\textwidth]{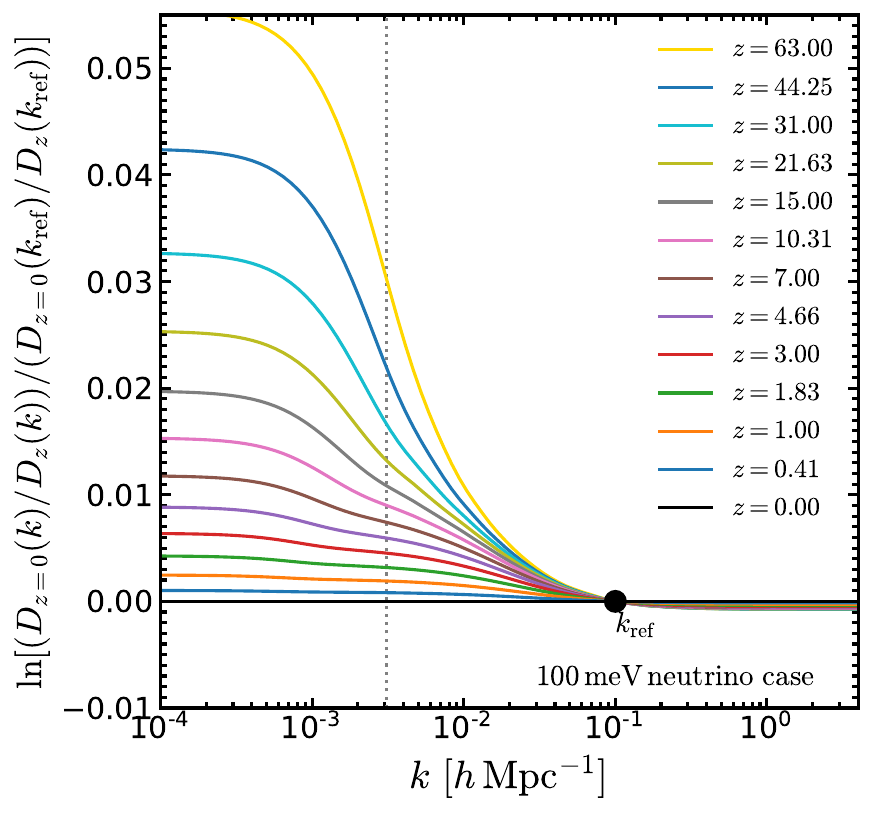}
\caption{Evolution of the scale dependence of the linear growth factor for cases with massless (left panel) and with massive neutrinos (right panel). The coloured lines correspond to different redshifts from $z=63$ to $z=0$, as labelled. Each of the lines shows the linear growth factor for the corresponding redshift as a function of wavenumber, normalised to the value of the growth factor at a fiducial reference wavenumber of $k_{\rm ref} = 0.1\,h\,{\rm Mpc}^{-1}$. The vertical dotted line indicates the fundamental mode of our largest simulation box with side-length $3000\,{\rm Mpc}$. The scale dependence around this mode stems from the difference between the evolution of super- and sub-horizon modes and is not captured properly by simulation codes that use Newtonian gravity. For massless neutrinos, the linear growth factor is however effectively scale independent for $k>k_{\rm ref}$. Even this property is lost for massive neutrinos due to their transitioning from being relativistic to being non-relativistic over this redshift range.}
\label{fig:Dk}
\end{figure*}

In terms of present day values, the contribution of the relativistic species, $\Omega_r(z)$ can be written as,
\begin{align}
\Omega_r(z) &= \Omega_{\gamma}(z) + \Omega^{\rm rel}_{\nu}(z)\nonumber\\ 
&= \Omega_{\gamma_0} \left[1 + N^{\rm rel}_\nu \frac{7}{8} \left(\frac{T_\nu}{T_{\rm CMB}}\right) \right](1+z)^4\left(\frac{H_0}{H(z)}\right)^2\label{eq:Or}\,,
\end{align}
where the present day value of the photon density parameter is
\begin{equation}\label{eq:Og0}
\Omega_{\gamma_0} = \frac{\pi^2}{15} \frac{(k_{\rm B} T_{\rm CMB})^4}{\hbar^3 c^5} \frac{8\pi G}{3H^2_0}\,,
\end{equation}
$N^{\rm rel}_\nu$ is the number of relativistic neutrino species, $k_{\rm B}$ is the Boltzmann's constant, $T_\nu$ $(1.9524\,{\rm K})$ and $T_{\rm CMB}$ $(2.7255\,{\rm K})$ are the neutrino and CMB photons temperature at the present time\footnote{The neutrinos and photons temperature are linked via the relation $T_\nu/T_{\rm CMB} = (4/11)^{1/3}$, assuming instantaneous decoupling. If one wants to take into account distortions induced by non-instantaneous decoupling and neutrino flavour oscillations, one usually addresses this by a slightly modified value of $N_\nu^{\rm rel}$.},  $H_0$ is the present day value of the Hubble parameter, and $G$ is the gravitational constant. On the other hand, the cold matter density parameter, $\Omega_{\rm{cb}}(z)$, is
\begin{equation}
\Omega_{\rm{cb}}(z) = \Omega_{\rm{cdm}}(z) + \Omega_{\rm{b}}(z) =(\Omega_{\rm{cdm}_0} + \Omega_{\rm{b}_0})(1+z)^3 \left(\frac{H_0}{H(z)}\right)^2\,.
\end{equation}
The massive neutrinos contribution in Eq.~\eqref{eq:O_nu}, is given by,
\begin{align}
\Omega_\nu(z) =& \frac{15}{\pi^4} \left(\frac{T_\nu}{T_{\rm CMB}}\right)^4 N_\nu \Omega_{\gamma_0} (1+z)^4\nonumber \\
&\times \mathcal{F} \left[\frac{M_\nu/(N_\nu k_{\rm B} T_\nu)}{1+z}\right]\left(\frac{H_0}{H(z)}\right)^2 \label{eq:O_nu}\,,
\end{align}
where $\Omega_{\gamma_0}$ is given by Eq.~(\ref{eq:Og0}), $N_\nu$ is the number of non-relativistic massive neutrinos, $M_\nu = \sum_i m_{\nu,i}$ is the sum of the masses of the degenerate massive neutrinos and the function $\mathcal{F}$ is defined as,
\begin{equation}
 \mathcal{F}(y) \equiv \int^{\infty}_0 \frac{x^2\sqrt{x^2 +
     y^2}}{1+e^x}{\rm d}x\,.
\end{equation}
The total matter contribution to the energy density is,
\begin{equation}
\Omega_{\rm m}(z) = \Omega_{\rm cb}(z) + \Omega_\nu(z)\,,   
\end{equation}
and the neutrino mass fraction today, $f_\nu$, can be written as,
\begin{equation}
    f_\nu(z) = \frac{\Omega_\nu(z)}{\Omega_{\rm m}(z)}\,.
\end{equation}
Finally, the Hubble rate throughout cosmic time in a flat Universe is given by
\begin{equation}\label{eq:H_nu}
H(z) = H_0[\Omega_{r0} (1+z)^4 + \Omega_{\rm{cb}0}(1+z)^3 + \Omega_\nu E^2(z) + \Omega_\Lambda]^{1/2}\,,
\end{equation}
where $E(z)\equiv H(z)/H_0$.

The evolution of the density parameters of the different matter and energy components, $\Omega_{\rm X}(a)$, as a function of the scale factor $a$ is shown in Figure~\ref{fig:Omegas}. In the displayed case, we have considered a relativistic massless neutrino $(N^{\rm rel}_\nu = 1)$ and two degenerate massive neutrinos $(N_\nu = 2)$ with masses $m_{\nu,i} = 50$ meV each, i.e., $M_\nu =\sum_i m_{\nu,i} = 100$ meV. This also corresponds to the case we used in our largest simulations with massive neutrinos.

\subsection{Initial conditions and linear growth with massive neutrinos}
\label{sec:ICs}
Traditionally, N-body simulations have been evolved using Newtonian gravity in the context of a homogeneous Friedmann-Lemaitre background solution \citep{Angulo2022}, although a few authors have included relativistic effects on large scales \citep[e.g][]{Adamek2016, East2018}. This is justified because once large-scale perturbative modes are inside the horizon, a Newtonian treatment accurately recovers their linear evolution. At the same time, non-linear evolution occurs on small enough scales and involves weak enough gravitational fields that general relativistic corrections are negligible, except in the immediate vicinity of black holes and neutron stars.

However, when simulation boxes with sizes of around $1 \Gpch$ and beyond are considered, the largest modes are actually of super-horizon size, at least at high redshift. Then ordinary Newtonian N-body codes do not deliver a correct growth rate on these scales, and questions related to gauge choices arise as well \citep[e.g.][]{Montandon2024}. In simulations without neutrinos, this is usually not a significant concern, as one typically constructs the initial conditions by back-scaling the linear theory power spectrum predicted by a Boltzmann code at $z=0$ to the initial redshift, using a $k$-independent linear theory growth factor. This yields an initial power spectrum that is not fully correct at the starting redshift, but the Newtonian code will evolve it on these large scales to arrive at the correct state at low-redshift.

When massive neutrinos are included, the same trick no longer works, because the neutrinos themselves introduce a scale-dependence in the linear growth factor. Furthermore, this linear theory growth factor can no longer be computed through a closed-form integral, unlike for flat cosmologies that feature only cold matter and dark energy \citep{Linder2003}. Different approaches have been proposed in the literature to deal with this complication when constructing initial conditions, including sophisticated rescaling techniques, as realised, for example, in the {\small REPS} code \citep{Zennaro2017, Zennaro2019}. These are needed, in particular, if the linear evolution of an N-body code does not exactly reproduce the linear evolution of a Boltzmann code, even on small scales, due to the (very common) omission of a consistent accounting of relativistic components such as the background photon field.

Our strategy has been to eliminate this latter error by outfitting {\small GADGET-4} both with a treatment of relativistic components in addition to cold matter, and with a consistent computation of the Hubble rate in the presence both of radiation components (photons and massless neutrinos), and of massive neutrinos, accounting also for the gradual transition of the latter from the relativistic to the non-relativistic regime. We can then directly use the transfer functions computed by a Boltzmann code such as {\small CAMB} \citep{Lewis2000} or {\small CLASS} \citep{Lesgourgues2011} at the starting redshift, i.e.~without requiring any back-scaling, because now the simulation code correctly accounts for the linear growth, as long as both the initial mode amplitudes and their initial growth rates are initialised correctly. The latter requires the logarithmic growth rate for every mode at the starting redshift. We obtain this from the Boltzmann code by numerically differentiating the growth factor based on two transfer function outputs spaced equidistantly around the desired starting redshift. This is done as a function of wave number  $k$, and for dark matter and massive neutrinos separately. 

This still leaves the problem of super-horizon modes. If we, for example, would simply initialise our simulation with the linear theory power spectrum as computed by {\small CAMB} at our starting redshift $z=63$, then the fundamental mode in our large box ($L = 2040 \Mpch = 3000\,{\rm Mpc}$) would lag behind in growth by about 2\% by redshift zero in a case without neutrinos, whereas in a $M_\nu=100$ meV model this lag would exceed 3\%. This effect is illustrated in Figure~\ref{fig:Dk}, and it arises because Newtonian gravity underpredicts the growth of these super-horizon modes in the matter dominated regime. Note that the effect gets even larger for still bigger simulation boxes, while becoming negligibly small at late times and for small boxes. In any case, we approach this by multiplying the initial power spectrum with a correction factor that boosts the linear theory input power spectrum such that  at $z=0$, after evolution with the Newtonian code, it will coincide with the result of the Boltzmann code \citep[see Fig.~3 of ][for corresponding measurements of the large-scale power spectrum both at the starting redshift and at $z=0$ for our MTNG3000 simulation, illustrating this approach]{Hernandez-Aguayo2023}. Note that the power spectrum will also agree with the Boltzmann code at all intermediate redshifts in the presence of massive neutrinos, unlike for the standard rescaling approaches. Only at Gpc-scales and beyond, the amplitudes can differ at the percent level due to the super-horizon correction, as illustrated in Figure~\ref{fig:Dk}.

We have implemented the above corrections in the version of our initial conditions code  {\small N-GENIC} built into {\small GADGET-4}. It uses different transfer functions as computed by the  Boltzmann code {\small CAMB} for cold dark matter and massive neutrinos. In addition, we use the Boltzmann code to compute derivatives of the corresponding transfer functions with respect to the scale factor, as needed to initialise the velocities of the matter components. Finally, we apply these through $2^{\rm nd}$-order Lagrangian perturbation theory to the unperturbed particle load, here modelled as two Cartesian grids for cold dark matter and neutrinos. The dark matter particles were aligned with the $10240^3$ Fourier grid used to compute the displacement field, so that no interpolation to the particle positions was necessary. The unperturbed neutrino grid was shifted relative to the cold dark matter particles by half a neutrino grid spacing. Note, however, that the large streaming velocities of the neutrinos quickly decorrelates their coordinates relative to the dark matter particles anyway. As the resolution of the neutrino grid is much lower than the dark matter grid, the Fourier grid substantially over-resolves the Nyquist frequency of the neutrinos, so that interpolation errors of the displacements to the neutrino positions are likewise negligible.

\subsection{The $\delta f$-method}
\label{sec:delta-f}
Recently, \citet{Elbers2021} proposed a novel approach to represent massive neutrinos in cosmological simulations, which has become known as the $\delta f$-method. We here provide our own derivation of this technique.

Let $f(\vec{x}, \vec{v}, t)$ be the neutrino distribution function, i.e.~the number density of neutrinos of mass $m_\nu$ as a function of phase space coordinates (expressed in terms of comoving coordinates and conjugate velocities) and time. Then we can obtain the neutrino mass density, 
\begin{equation}
  \rho(\vec{x}, t)  = m_{\nu} \int f(\vec{x}, \vec{v}, t) {\textrm{
      d}}\vec{v} ,
\end{equation}
by integrating out the velocity distribution. Now let us introduce the time evolution of a spatially undisturbed neutrino background, $f_0 = f_0(\vec{v}, t)$, and consider the difference between the actual neutrino density and this background:
\begin{equation}
  \rho(\vec{x}, t)  = m_{\nu} \int f_0  {\textrm{ d}}\vec{v}  +  m_{\nu}  \int\left[
    f(\vec{x}, \vec{v}, t)  - f_0  \right] {\textrm{ d}}\vec{v}.
\end{equation}
Let us now integrate over some finite volume $V$ (which could be a computational cell, or some region of the simulation box) in order to simplify the argument we are going to make later on. The enclosed neutrino mass inside in this volume is then
\begin{equation}
  M_{\nu} (t)  =\rho_0 V  +  m_{\nu}  \int\left[
    f(\vec{x}, \vec{v}, t)  - f_0  \right] {\textrm{ d}}\vec{v} {\textrm{ d}}\vec{x},
\end{equation}
where $\rho_0$ stands for the (time-dependent) average density of neutrinos. We now Monte-Carlo integrate the integral on the {\it rhs} with a sample of $N$ points in the chosen volume. We  use a sampling function $f^*$ proportional to $f$ itself (which is the concept of importance sampling), with $f^*$ normalised  as $f^* = (1 / N_{\nu}) f$, where $N_{\nu}$ is the number of neutrino particles in the volume.
We can thus write the integral as:
\begin{equation}
  M_{\nu} (t)  =\rho_0 V  + m_{\nu}  \int\frac{
    f - f_0  }{ f^*}\;  f^*  {\textrm{ d}}\vec{v} {\textrm{ d}}\vec{x}.
\end{equation}
By construction, ${\textrm{ d}}p = f^*{\textrm{ d}}\vec{v} {\textrm{ d}}\vec{x}$ is the probability to find a sampling point in an infinitesimal phase-space volume around $({\textrm{d}}\vec{x}, {\textrm{d}}\vec{v})$.
The integral can therefore be approximated as
\begin{equation}
  \int\frac{
    f - f_0  }{ f^*} \; f^*  {\textrm{ d}}\vec{v} {\textrm{ d}}\vec{x}
 \; \simeq  \;\frac{1}{N} \sum_i \frac{ f(\vec{x}_i, \vec{v}_i, t)  -
  f_0(\vec{v}_i, t) }{ f^*(\vec{x}_i, \vec{v}_i, t)} .
\end{equation}
A key step is now to argue that the sampling points need not necessarily be drawn from the current $f$, but instead can be identified with sampling points used to sample the initial conditions at the initial time $t_0$, and which were evolved along characteristics to the current time $t$.

At this initial time, let us assume that the sampling points have phase-space coordinates $(\vec{x}_i^0, \vec{v}_i^0)$. The evolution of the particles between $t_0$ and the current time $t$ is governed by the collisionless Boltzmann equation, which preserves the fine-grained phase-space density along individual particle orbits. We hence know that we have  $f(\vec{x}_i, \vec{v}_i, t) = f(\vec{x}_i^0, \vec{v}_i^0, t_0)$. We therefore can write:
\begin{equation}
  M_{\nu} (t)  =\rho_0 V  + 
\frac{m_{\nu} N_{\nu}}{N} \sum_i \left[ 1 - \frac{ f_0(\vec{v}_i, t) } { f(\vec{x}_i^0, \vec{v}_i^0, t_0)}\right].
\end{equation} 
The prefactor $m^*_{\nu} \equiv m_{\nu} N_{\nu} / {N}$ is the fiducial simulation particle mass we use to represent the neutrinos. The quantity $f_0(\vec{v}_i, t)$ is known and can simply be evaluated for the current velocity of the particle as the background density times the Fermi-Dirac distribution $F$  evaluated for $\vec{v}_i$:
\begin{equation}
  f_0(\vec{v}_i, t) = \rho_0 F(\vec{v}_i)\,.
\end{equation}
A further assumption we make is that the initial phase-space distribution can be written as
\begin{equation}
f(\vec{x}_i^0, \vec{v}_i^0, t_0) \simeq \rho_0 (1+\delta_i^0) F(\vec{v}_i^0),
\end{equation} 
where $\delta_i^0$ is the initial density perturbation for the neutrino particle~$i$. This approximation is not exact and thus will introduce some initial transients that take some time to decay away. See \citet{Elbers2022} for ways to improve this further, the simulations we consider here are however run with this physical approximation. Using it, the $\delta f$-method predicts the neutrino mass in our test volume as
\begin{equation}
  M_{\nu} (t)  =\rho_0 V  + 
m^*_{\nu} \sum_{\substack{i \\ \vec{x}_i \in V}} \left[ 1 - \frac{ F(\vec{v}_i) } {(1 +\delta_i^0)  F(\vec{v}_i^0)}\right],
\end{equation}
which is equivalent to assigning each neutrino particle a variable mass
\begin{equation}
\tilde{m} = m^*_{\nu} \left[ 1 - \frac{
      F(\vec{v}_i) } {(1 +\delta_i^0)  F(\vec{v}_i^0)}\right],
\end{equation}
based on its current phase-space coordinates. This mass, instead of $m^*_{\nu}$, is used for computing the gravitational field with the Poisson solver implemented in the {\small GADGET-4} code. This importance sampling greatly reduces sampling noise in the neutrino density field, a feature that is made possible by exploiting not only the current coordinate of the simulation particles, but also their current velocities. Note however that this does not change the character of each particle as representing a random neutrino orbit. All particles still follow an unchanged equation of motion, except that the force they experience is less affected by shot noise.

In our discussion so far we have ignored the fact that the neutrinos may have relativistic velocities, particularly at early times or for light neutrino masses. We address this by using the particle momentum $\vec{p}_\gamma = m_\nu \vec{v} /( 1-v^2/c^2)^{1/2}$ in the above phase space arguments, and not the kinematic velocity $\vec{v}$ itself. Note that it is also the momentum that we obtain from the initial Fermi-Dirac distribution function. We integrate the neutrino particle momenta in the equations of motion by  approximating their temporal change with the ordinary Newtonian gravitational force, with neutrinos contributing with their rest masses to the source function of the Poisson equation. The particles are moved in space, however, with their kinematic velocity $\vec{v}$. Note that the expansion of space also redshifts $\vec{p}_\gamma$, and in that way automatically induces the correct transition from the relativistic to the non-relativistic regime.

\begin{figure}
 \centering
\resizebox{8.3cm}{7cm}{\includegraphics{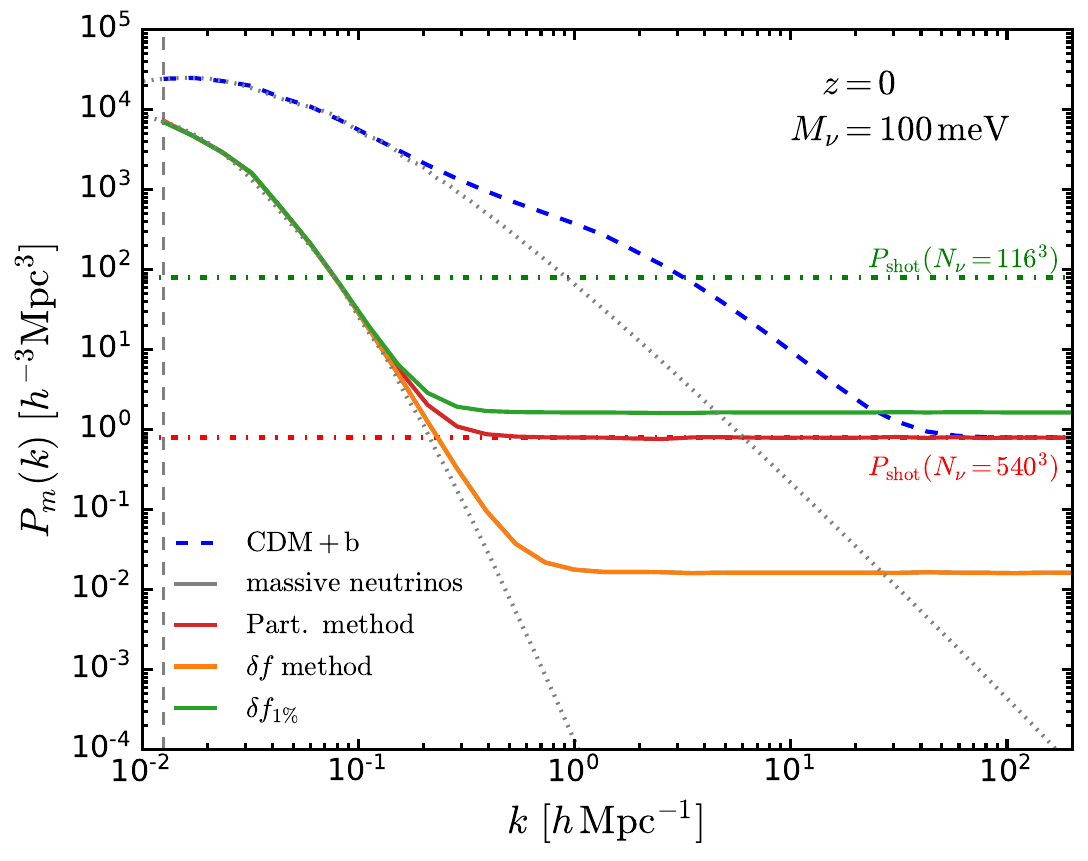}}
\caption{Matter power spectra at $z=0$ in test runs with a $L=500\Mpch$ box and $N_{\rm cdm} = 540^3$ cold dark matter particles, using two different neutrino modelling methods. We show results with $N_\nu = 540^3$ neutrino particles both for an ordinary particle method (red solid line) and for the $\delta f$-method (orange line). We also illustrate the impact of using only 1\% of this neutrino particle number (equivalent to $N_\nu = 116^3$) in the $\delta f$-method (green solid line). The dashed blue curve gives a measurement of the power spectrum of the CDM plus baryon component of the simulation, for comparison. The horizontal green and red dashed-dotted lines indicate the expected shot-noise levels for Poisson samples with $116^3$ and $540^3$ particles, respectively. The $\delta f$-method is able to estimate the neutrino density field with substantially sub-Poissonian discreteness noise.}
  \label{fig:Pk_test}
\end{figure}

\begin{figure*}
 \centering
\includegraphics[width=1.0\textwidth]{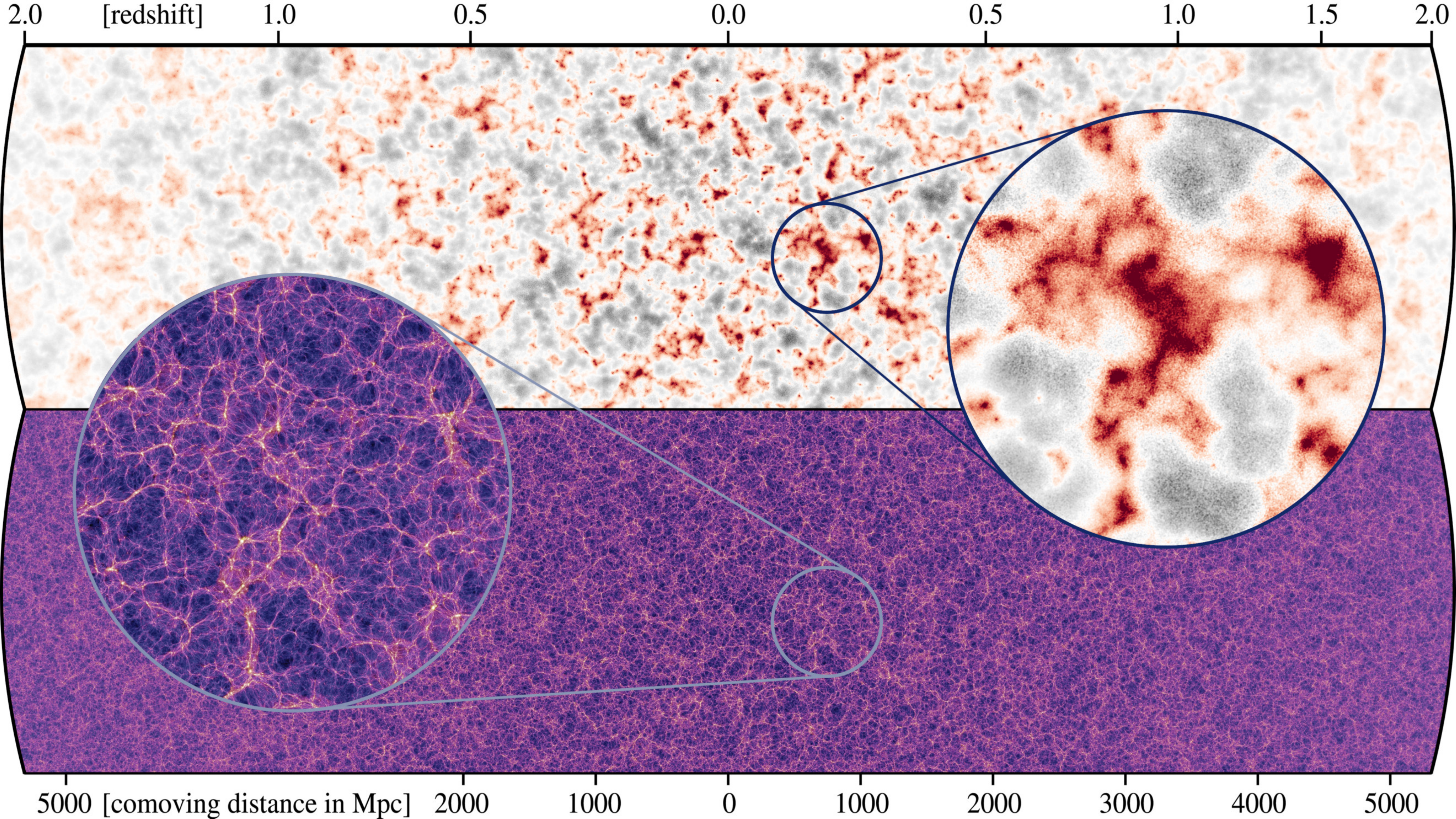}
\caption{Massive neutrinos (top map) and cold dark matter (bottom map) distributions on the past lightcone of a fiducial observer positioned at the centre of the two horizontal stripes. As cosmic expansion slows down the neutrinos at late times (corresponding to small redshift and distance), they start to cluster around the biggest concentrations of dark matter as shown by a comparison of the zoomed insets. This slightly increases the mass and further growth rate of these largest structures. These images are of a rectangular subarea of one of the disc-like lightcones described as Cone 3 in Sec.~3.5 of \citealt{Hernandez-Aguayo2023}. In particular, the projection depth is $15\Mpch = 22.14\,{\rm Mpc}$. The two nested zoom-in regions have a diameter of 400 Mpc.}
  \label{fig:NeutrinoLightCone}
\end{figure*}

As \citet{Elbers2021} first demonstrated, the $\delta f$-method is indeed capable of reducing the shot noise of the neutrino particle distribution in cosmological N-body simulations. We show an example of this based on our own simulations in Fig.~\ref{fig:Pk_test}, where we compare the total matter power spectrum of three neutrino simulations for the $M_\nu = 100$ meV model at the present time, using either the plain particle method or  the $\delta f$-method. Our test simulations consist of boxes with side-length of $500\,\Mpch$ and use $540^3$ CDM particles. For the particle method we addionally evolve $540^3$ neutrino particles, and for the $\delta f$-method we show two configurations with $540^3$ and $116^3$ extra neutrino particles, respectively\footnote{Note that $116^3$ is roughly 1\% of $540^3$.}.

When comparing the power spectra measurements for the simulations with $540^3$ neutrino particles (red and orange solid lines), we can see from Fig.~\ref{fig:Pk_test} that the $\delta f$-method reduces the shot-noise of the neutrino distribution by almost two orders of magnitude. In addition, we find that using only $116^3$ neutrino particles with the $\delta f$-method (green solid line) produces nearly the same level of shot-noise as using $540^3$ neutrino particles with the particle method (red solid line). This is informative for selecting the relative particle numbers of neutrinos and cold dark matter in large-scale cosmological simulations with massive neutrinos. For the calculations we present later in this paper, the neutrino particle number is chosen such that the resulting shot noise level of the neutrinos at $z=0$ is very similar to that of the cold dark matter particles \citep[see also][]{Hernandez-Aguayo2023}.

\section{Simulations with massive neutrinos}
\label{sec:mtng}
We here focus on the neutrino simulations of the MillenniumTNG project \citep[MTNG;][]{Hernandez-Aguayo2023,Pakmor2023} which presently consist of two large-volume cosmological simulations with box size $L=2040\Mpch$ (3 Gpc; MTNG3000) and massive neutrinos with a summed mass of $M_\nu = 0.1$ eV, in addition to several $430\Mpch$ (632.4 Mpc; MTNG630) runs with different neutrino masses. All the simulations were run with the {\small GADGET-4} simulation code \citep{Springel2021} using the $\delta f$-method outlined above (cf. Sec.~\ref{sec:delta-f}). For each of the models, we have run two realisations using the fixed-and-paired technique of \citet{Angulo:2016hjd}, which we denote in the following as ``A'' and ``B'' runs. The initial conditions of the runs of each pair differ in the phases of each perturbation mode by 180 degrees, or in other words, the sign of the linear density fluctuations is reversed. As \citet{Angulo:2016hjd} have shown, by averaging the results of such a simulation pair, certain leading higher-order deviations from linear theory cancel, effectively suppressing the variance compared to a random realisation in the same volume, and thus leading to an improved estimate of the expected cosmological mean.

Substantial computational resources, particularly memory, are needed to generate the very large MTNG3000 initial conditions.  To retain the option of being able to make relatively inexpensive zoom initial conditions for objects in these volumes we use the Panphasia method outlined in \citet{Jenkins2013} to set the phases for the two MTNG3000 simulations. The essence of this method is to construct the Gaussian white noise field, from which the phases of the simulation are inherited, in a hierarchical way based on an octree spatial structure. The white noise field is built from piecewise polynomials localised within octree cells. Coupling this construction method with a suitable pseudo-random number generator it is possible to define extremely large realisations of the phases that cover all physical scales of interest for the original simulations and any follow up zoom simulations. The hierarchical structure of the field lends itself to the efficient creation of zoom initial conditions.

This approach, however, is not compatible with the {\it fixing} part of the fixed-and-paired method \citep{Angulo:2016hjd} when applied to all Fourier modes. In practice the variance suppression we need can be achieved by fixing just the poorly sampled low wavenumber Fourier modes as the sheer number of high-$k$ modes guarantees behaviour close to the mean.  For MTNG3000, only Fourier modes with wavelengths greater or equal to 50~Mpc were fixed -- a total of 452044 independent modes -- while all other mode amplitudes are left free. With almost half a million fixed modes the measured linear power spectrum of the free modes is within 1\% of the target for bins at least as wide as the fundamental mode of the periodic volume. 

The fixed modes are confined in the Fourier domain to large wavelengths and it must be possible to represent these fixed modes very accurately using polynomials confined to the largest octree cells. If that is the case the information in the smaller octree cells can be treated as independent allowing the creation of efficient zoom initial  conditions. This can be achieved by using higher degree polynomials than was used in the original Panphasia  field \citep{Jenkins2013}, which was built from just degree one polynomials.  Fortunately the computational cost of going to higher degree polynomials can be offset in practice by much more efficient code for computing the Panphasia field and a six degree scheme has already been implemented in the public {\small MONOFONIC} cosmological initial conditions code \citep{Hahn2020, Michaux2021}.  This version of the Panphasia field has also been used for the {\small FLAMINGO} project \citep{Schaye2023}. As with the original Panphasia field the phases for the MTNG3000-A volume are described by a plain text `phase descriptor', which in this case contains additional information about the degree of the polynomials and which Fourier modes are fixed. We reproduce here it for future reference\footnote{[Panph6, L19, (301309,431832,75016), S5, KK3600, CH2926231125, MillenniumTNG]}.

The fiducial cosmology of our $100\,{\rm meV}$ neutrino simulations is consistent with the DES-Y3 constraints \citep{DES:2021wwk}: $\Omega_{\rm cb} = 0.30368$, $\Omega_{\rm b} = 0.04870$, $\Omega_{\Lambda} = 0.69393$, $h = 0.68$ and $\sigma_8 = 0.8040$. In addition, two degenerate neutrino species each with 50 meV rest mass, yielding $\Omega_\nu = 2.324\times 10^{-3}$, one massless neutrino species ($\Omega_\nu^{\rm rel} = 1.232\times 10^{-5}$), and a standard photon background accounting for the cosmic microwave background with $T_{\rm CMB}=2.7255\,{\rm K}$ were assumed ($\Omega_{\gamma} = 5.348\times 10^{-5}$). This means that $0.76$~percent of today's mean matter density are made up of neutrinos in this cosmology -- roughly half of what the stellar mass contributes.

The ICs were generated at $z=63$ with the method described in Sec.~\ref{sec:ICs} using the linear-theory transfer functions computed by {\small CAMB}. The MTNG3000 runs follow close to 1.1 trillion ($10240^3$) cold dark matter and more than 16 billion ($2560^3$) massive neutrino particles\footnote{Note that thanks to the $\delta f$-method (cf. Sec.~\ref{sec:delta-f}), we need only 1.5 percent of the  CDM particle number as neutrino particles in order to achieve a similar shot noise level.}, leading to a cold dark matter particle mass resolution of $6.66\times10^8\Msh$, which is more than twice the resolution of the original Millennium simulation \citep{Springel2005}, but in a much larger volume, and making them some of the largest and highest-resolution simulations with massive neutrinos run to date. The MTNG630 runs evolve $2160^3$ cold dark matter and $540^3$ massive neutrino resolution elements, giving the same particle mass resolution as MTNG3000.

In addition, we have run  $M_\nu = 0\,{\rm eV}$ (massless case) and $M_\nu = 0.3\,{\rm eV}$ models for the MTNG630 runs only. In these cases, we changed the $\Omega_{\rm cb}$ cosmological parameter in order to keep the value of the total matter density $\Omega_{\rm m} = \Omega_{\rm cb} + \Omega_{\nu} = 0.306$ fixed when the massive neutrino contribution is varied. We also slightly adjusted the value of $\Omega_\Lambda$ in order to compensate for the changing relativistic neutrino contribution and to retain an  exactly flat cosmology. For the normalisation of the initial power spectrum, we kept the CMB normalisation on the largest scales invariant, implying small changes in the linearly extrapolated small-scale $\sigma_8$ value at $z=0$.

\begin{table*}
\begin{tabular}{cccccc}
\hline
Model   & $\Omega_{\rm m}$ & $\Omega_{\rm cdm}$ & $\Omega_{\rm b}$ & $\Omega_\nu$ & $\sigma_8$ \\ \hline
0 eV    & 0.3060           & 0.2573             & 0.0487           & 0.0          & 0.8230     \\
0.1 eV  & 0.3060           & 0.254976           & 0.0487           & 0.002324     & 0.8040     \\
0.3 eV  & 0.3060           & 0.250329           & 0.0487           & 0.006971     & 0.7623     \\
MTNG740 & 0.3089           & 0.2603             & 0.0486           & $-$          & 0.8159     \\ \hline
\end{tabular}
\caption{Values of the cosmological matter parameters at $z=0$ for the MTNG630 neutrino simulations and the MTNG740 neutrino-less simulation.}
\label{tab:param}
\end{table*}

\begin{table*}
\centering
\begin{tabular}{ccccccccccc}
\hline
Type    & Run name        & Box size  & $N_{\rm cdm}$ & $N_{\nu}$ & $N_{\rm gas}$ & \multicolumn{1}{c}{$m_{\rm cdm}$} & $m_{\nu}$ & $m_{\rm gas}$  & $\sum\,m_{\nu}$  & $\epsilon_{\rm cdm}$ \\
        &                       & $[\Mpch]$ &            &  &               & \multicolumn{1}{c}{$[\Msh]$}     & $[\Msh]$   & $[\Msh]$   &   $[{\rm eV}]$    & $[\kpch]$  \\ \hline
Neutrinos  & MTNG3000-DM-0.1$\nu$-A/B & 2040       & $10240^3$  & $2560^3$ &   $-$            & $6.66 \times 10^8$                &  $3.26 \times 10^8$    & $-$ & 0.1 & 4         \\
        & MTNG630-DM-0.3$\nu$-A/B      & 430       & $2160^3$  & $540^3$  &  $-$ & $6.54 \times 10^8$                & $9.76 \times 10^8$ &  $-$ & 0.3 & 4       \\  
        &  MTNG630-DM-0.1$\nu$-A/B      & 430       & $2160^3$  & $540^3$  &  $-$ & $6.66 \times 10^8$                & $3.26 \times 10^8$ & $-$ & 0.1 & 4       \\  
        &  MTNG630-DM-0.0$\nu$-A/B      & 430       & $2160^3$  & $-$  &  $-$ & $6.66 \times 10^8$                & $-$ & $-$ & 0.0  & 4       \\ \hline        
CDM only & MTNG740-DM-A/B    & 500       & $4320^3$   & $-$ &      $-$         & $1.32 \times 10^8$               &        $-$         & $-$ & $-$ & 2.5         \\ \hline
Full-hydro   & MTNG740-A      & 500       & $4320^3$   &   $-$ &  $4320^3$   & $1.12 \times 10^8$                & $-$ & $2.00 \times 10^7$ & $-$ & 2.5         \\ \hline
\end{tabular}
\caption{Numerical specification of the MillenniumTNG neutrino simulations with box sizes of $2040\Mpch$ and $430\Mpch$, with neutrinos of mass $\Sigma\,m_{\nu} = 100\,{\rm meV}$ (available for all box sizes), $\Sigma\,m_{\nu} = 300\,{\rm meV}$ and $\Sigma\,m_{\nu} = 0\,{\rm meV}$ (for MTNG630 runs only). In addition, we include the specifications of the MTNG740 and MTNG740-DM simulations, for comparison with the neutrino-less branch of the MTNG simulation project. }
\label{tab:sims}
\end{table*}

In Table~\ref{tab:param} we summarise the cosmological parameters of the simulations, i.e.~the corresponding values of the matter contributions (CDM, baryons and massive neutrinos) to the total density parameter, as well as the present-day values of the $\sigma_8$ parameter for each neutrino model. The simulations with varied neutrino mass help us to quantify the impact of massive neutrinos on galaxy observables such as the stellar mass function and the clustering of galaxies, as we will show below. The numerical specifications of our neutrino simulations are listed in Table~\ref{tab:sims}; here we also include the specifications of the MTNG740 runs without neutrinos (full-hydro and dark matter only) in the MillenniumTNG project, for completeness.

Following the same outputting strategy as for the MTNG740 simulations \citep{Hernandez-Aguayo2023}, we have measured the matter power spectra (total matter, cold matter and massive neutrinos) and produced halo/subhalo catalogues on-the-fly for 133 matching output times for all the neutrino simulations\footnote{Actually, for the MTNG3000-A run we have 13 additional output times.} between $0 < z < 63$  (with more than 120 halo and subhalo catalogues  at redshifts $z < 20$). Halo catalogues (FoF groups) were constructed using the friends-of-friends (FoF) algorithm \citep{Davis:1985} at all the output times, and subhaloes were identified with the {\small SUBFIND-HBT} algorithm \citep{Springel2001, Han2012} as implemented in the substructure finder of {\small GADGET-4} \citep[for more details see][]{Springel2021}. We also produced six particle lightcones for each run, keeping the same specifications and geometry as for the five lightcones of MTNG740 \citep[see Sec.~3.5 of][]{Hernandez-Aguayo2023}, but adding a sixth lightcone in the form of an additional pencil beam.

In Figure~\ref{fig:NeutrinoLightCone}, we show maps of the dark matter and neutrino density fields on one of the lightcones out to redshift $z=2$ \citep[see the description of Cone 3 in Sec. 3.5 of][]{Hernandez-Aguayo2023}. One can clearly see how the clustering strength of the dark matter gradually increases from $z=2$ at the left and right edges of the map to the observer in the middle of the slices  at $z=0$, whereas the neutrino density contrast is still very small at $z=2$ but then starts to respond to the largest dark matter structures that have formed in the low redshift universe due to the progressive redshifting of the thermal neutrino velocities. We also produced mass-shell maps with 1.8 billion pixels for weak-lensing studies \citep{Ferlito2023, Ferlito2024}. Furthermore, we have built merger trees from all subhaloes to allow the construction of physically motivated galaxy mock catalogues in massive neutrinos cosmologies, based on our updated version of the {\small L-GALAXIES} code for semi-analytic galaxy formation \citep{Barrera2023}.

\begin{figure*}
 \centering
\includegraphics[width=0.47\textwidth]{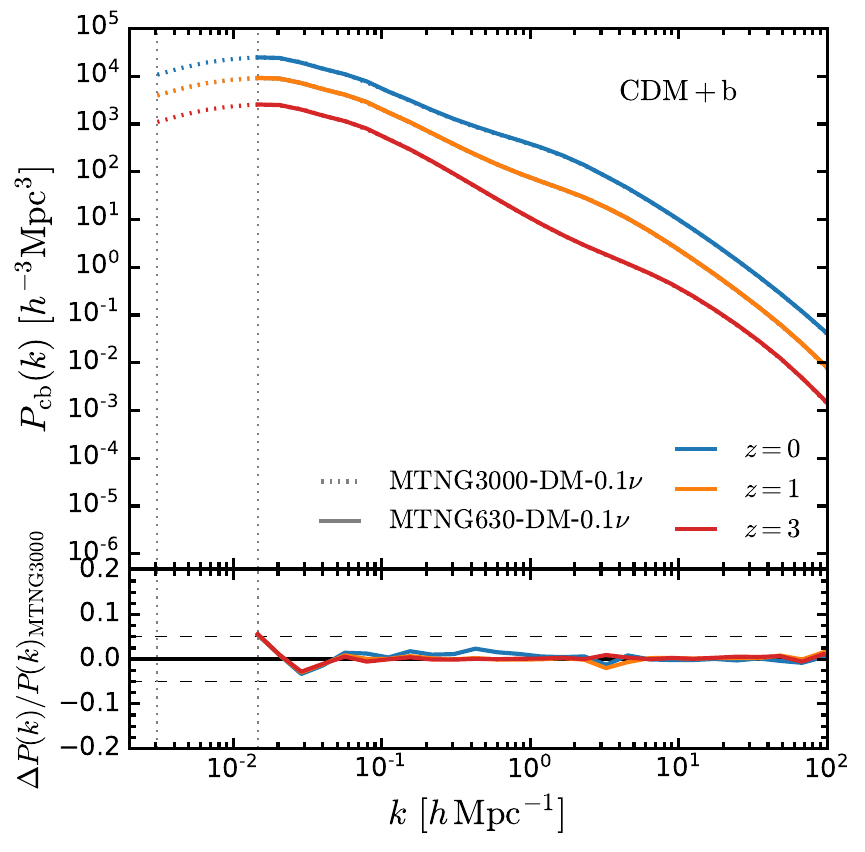}
\includegraphics[width=0.47\textwidth]{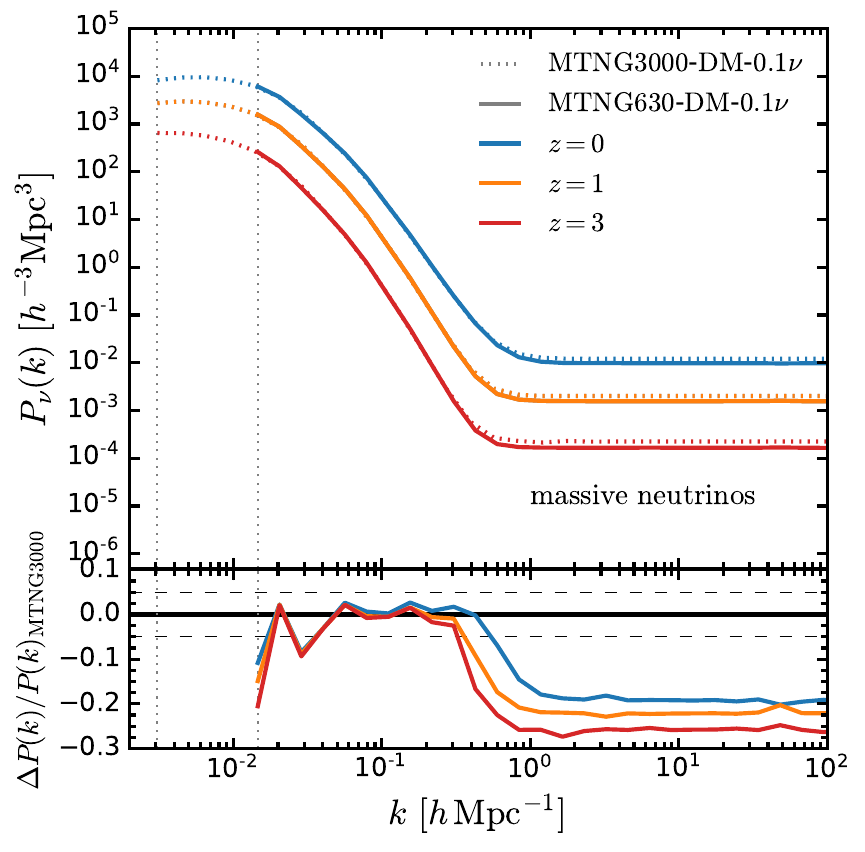}
\caption{Box size impact on the cold matter (CDM plus baryons; left panel) and massive neutrino (right panel) power spectra of our $100\,{\rm meV}$ MTNG simulations at $z=0$ (blue lines), $z=1$ (orange lines) and $z=3$ (red lines). Different line styles represent different simulations, as labelled. The vertical dashed lines indicate the fundamental modes of the two box sizes we used. Note that we are showing the average of the A and B realisations of the MTNG3000 and MTNG630 simulations. The lower subpanels give the fractional difference in power between the two box sizes with dashed horizontal lines marking a nominal 5\% difference.}
  \label{fig:Pk_box}
\end{figure*}

We note that the data size of the MTNG3000 runs is staggeringly large. Each of the simulations contains about $1.16 \times 10^9$  FoF haloes at $z=0$ (for a minimum of 32 particles per group), and $1.46 \times 10^9$ gravitationally bound subhaloes. There are about $\sim 1$ billion distinct merger trees\footnote{Subhaloes are stored in the same tree if they have a progenitor or descendant relation of any kind, or if they are part of the same FoF group at any time.} for each of the simulations, linking together about $1.5 \times 10^{11}$ subhaloes. Running {\small L-GALAXIES} on them produces about 1.4 billion galaxies in the volume of each of the simulations. A full particle snapshot of these simulations requires about 36 TB disc space, but thanks to the on-the-fly processing, we did not actually have to store this amount for the 133 output times (which would have led to a ballooning of the data volume to an inconvenient 4.7 PB -- just for one of the two large simulations).

\begin{figure*}
 \centering
\includegraphics[width=0.47\textwidth]{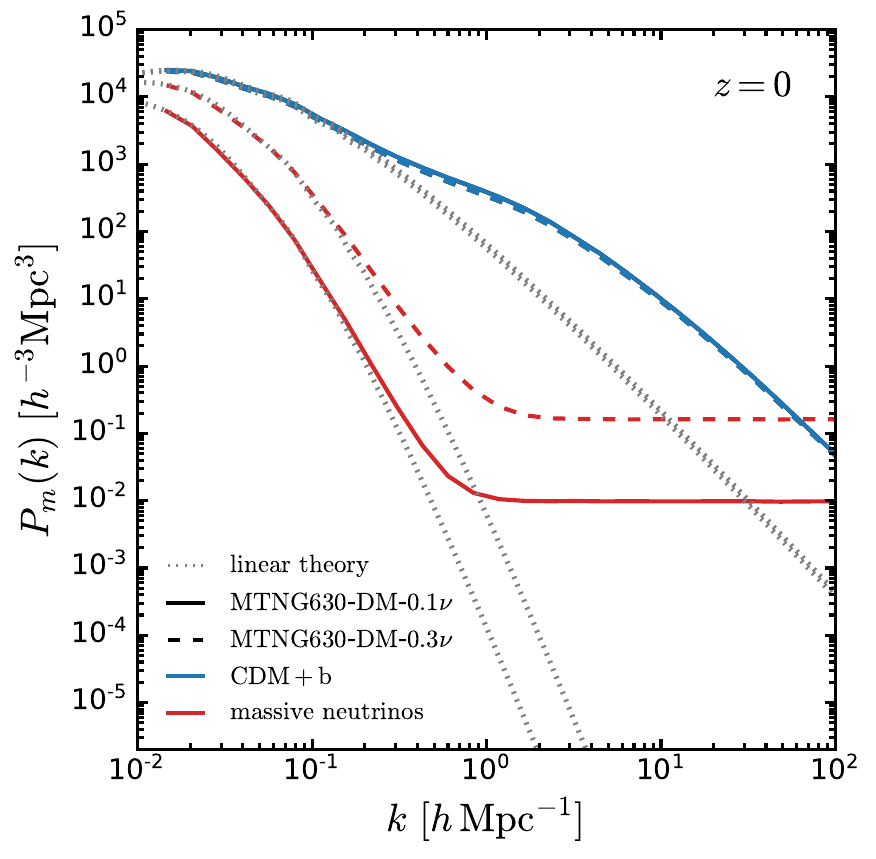}
\includegraphics[width=0.47\textwidth]{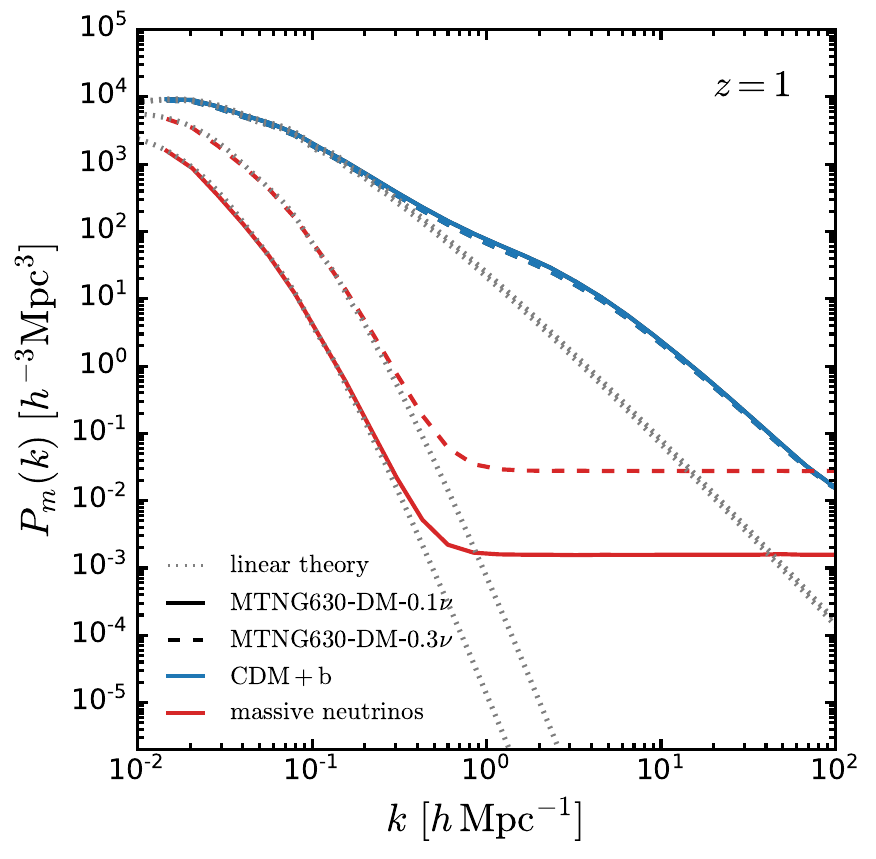}
\includegraphics[width=0.47\textwidth]{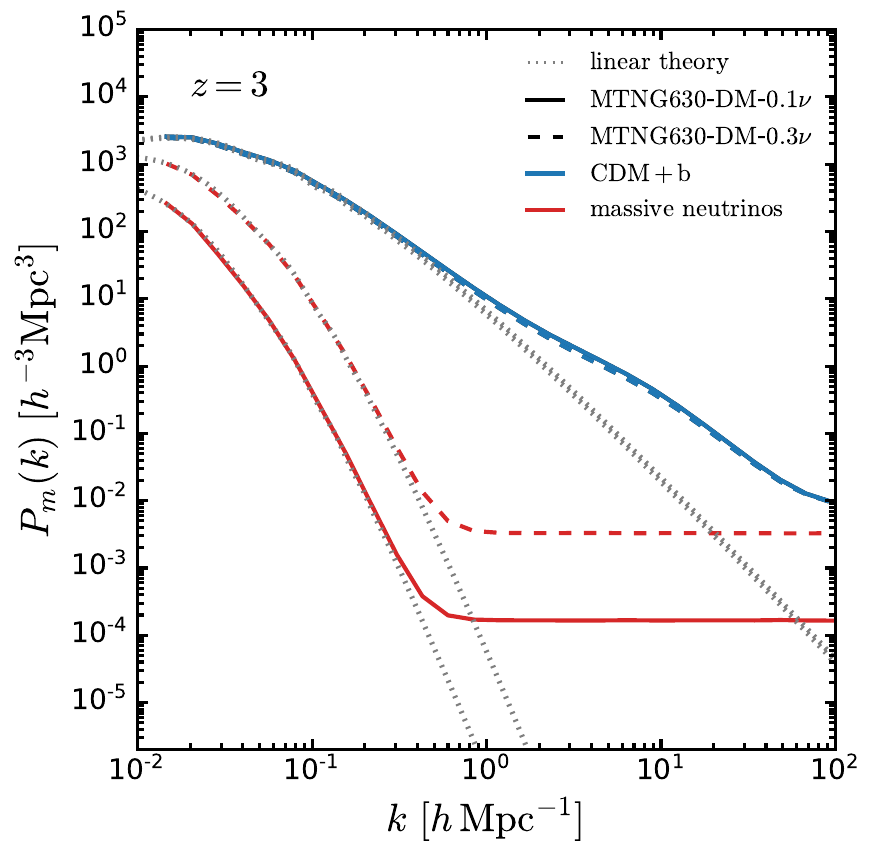}
\includegraphics[width=0.47\textwidth]{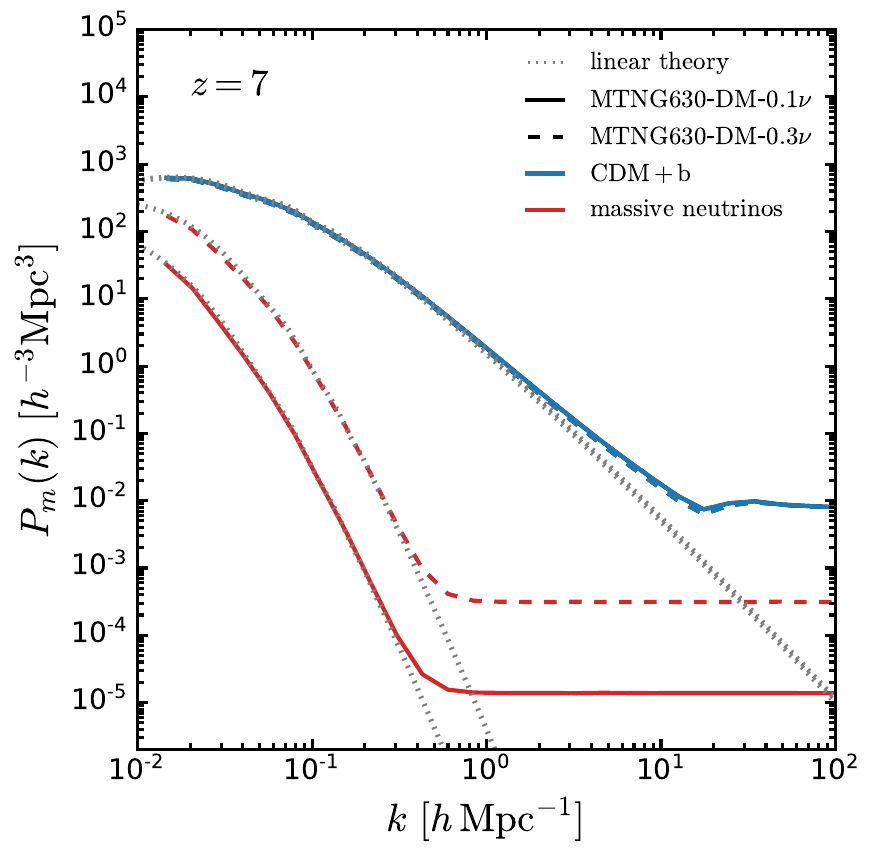}
\caption{Matter power spectra of the cold matter (CDM+b) and massive neutrinos components of the MTNG630-DM-$0.1\nu$ (solid lines) and MTNG630-DM-$0.3\nu$ (dashed lines) simulations at four different redshifts as indicated in each panel. Dotted curves represent linear theory predictions from {\small CAMB} at the corresponding times for each model. Note that we did not subtract any shot-noise contribution from the power spectrum measurements.}
  \label{fig:Pkm_L430}
\end{figure*}

\section{Matter clustering}
\label{sec:matter}
We show in Figure~\ref{fig:Pk_box} the clustering of the different matter components as measured through the power spectrum (separately for cold matter, $P_{\rm cb}(k)$, and massive neutrinos, $P_\nu(k)$) of our $100\,{\rm  meV}$ simulations (MTNG3000-DM-0.1$\nu$ and MTNG630-DM-0.1$\nu$) at redshifts $z=0$, $1$ and $3$. The vertical dotted lines indicate the fundamental mode of each box, $k_{\rm box} = 2\pi/L_{\rm box} \sim 0.0031\hMpc$ and $\sim 0.0146\hMpc$, respectively. 

Since the 100 meV neutrino cosmological model is the same for these two boxes, we can use the comparison to quantify how box size effects limit the accuracy of our simulations. In the lower left subpanel of Fig.~\ref{fig:Pk_box} we can see that the cold matter power spectra of the MTNG630 run agree almost perfectly with the MTNG3000 spectra, from large-scales $k\sim 0.03\hMpc$ down to very small-scales, $k \sim 100\hMpc$. Only small differences, less than 5 per cent, are seen between the small and the large box simulations near the fundamental wave number of MTNG630, reflecting only minor effects from the limited box size of this simulation. 

The box size impact on massive neutrino clustering, shown in the lower right subpanel of Fig.~\ref{fig:Pk_box}, is more noticeable. In this case, we find better than 2 per cent agreement between MTNG630 and MTNG3000 in the $k$-range $0.04 < k/[\hMpc] < 0.4$, which is the range over which the neutrino power spectrum is not dominated by shot noise. Interestingly, on still smaller scales -- in the noise dominated regime -- we observe a constant offset of $\sim$20 to 30 percent between MTNG630 and MTNG3000. In this regime, the power spectrum is below the nominal Poisson shot noise limit, thanks to the $\delta f$-method, but whereas the actual Poisson shot noise for the simulations is very similar ($P_{\rm shot}^{\rm Poisson}=L^3/N_\nu$), the much smaller effective noise level after using the $\delta f$-method shows a noticeable box size effect. This can be attributed to the stronger neutrino clustering in the large simulation box as a result of its much better representation of massive galaxy clusters and superclusters; this diminishes the noise-damping provided by the $\delta f$-method. The growth of such structures towards lower redshift is also the reason why the noise damping effect generally becomes weaker at late times.
In any case, these box-size tests indicate that the MTNG630 runs are still very helpful for cosmological analyses over the $k-$range of interest of current and future galaxy surveys such as DESI, Euclid and PFS, despite their limited box size.

Next we explore the time evolution of the clustering of the cold matter (CDM plus baryons) and massive neutrino components and compare the 100 meV and 300 meV neutrino models, using the MTNG630 simulations. The corresponding power spectra measurements are shown in Figure~\ref{fig:Pkm_L430} at redshifts $z=0$, $1$, $3$ and $7$. In all panels of Fig.~\ref{fig:Pkm_L430} there is excellent agreement on large scales between the simulation measurements and the linear theory predictions (which are computed directly with {\small CAMB} for the corresponding times) for both matter components and both neutrino models. There is a monotonic evolution in the increase of power from high redshift ($z=7$; lower right panel of Fig.~\ref{fig:Pkm_L430}) to the present day ($z=0$; upper left panel of Fig.~\ref{fig:Pkm_L430}), as  expected from linear theory.

We also observe that neutrinos in the more massive case ($M_\nu = 300\,{\rm  meV}$; see the red dashed lines in Fig.~\ref{fig:Pkm_L430}) are more clustered than the lighter neutrinos  ($M_\nu = 100\,{\rm meV}$; the red solid lines in Fig.~\ref{fig:Pkm_L430}). To compensate the higher neutrino clustering, note that the CDM+b component in the $300\,{\rm meV}$ model (blue dashed lines) ends up slightly less clustered than its $100\,{\rm  meV}$ (blue solid lines) counterpart.

\begin{figure*}
 \centering
\includegraphics[width=0.47\textwidth]{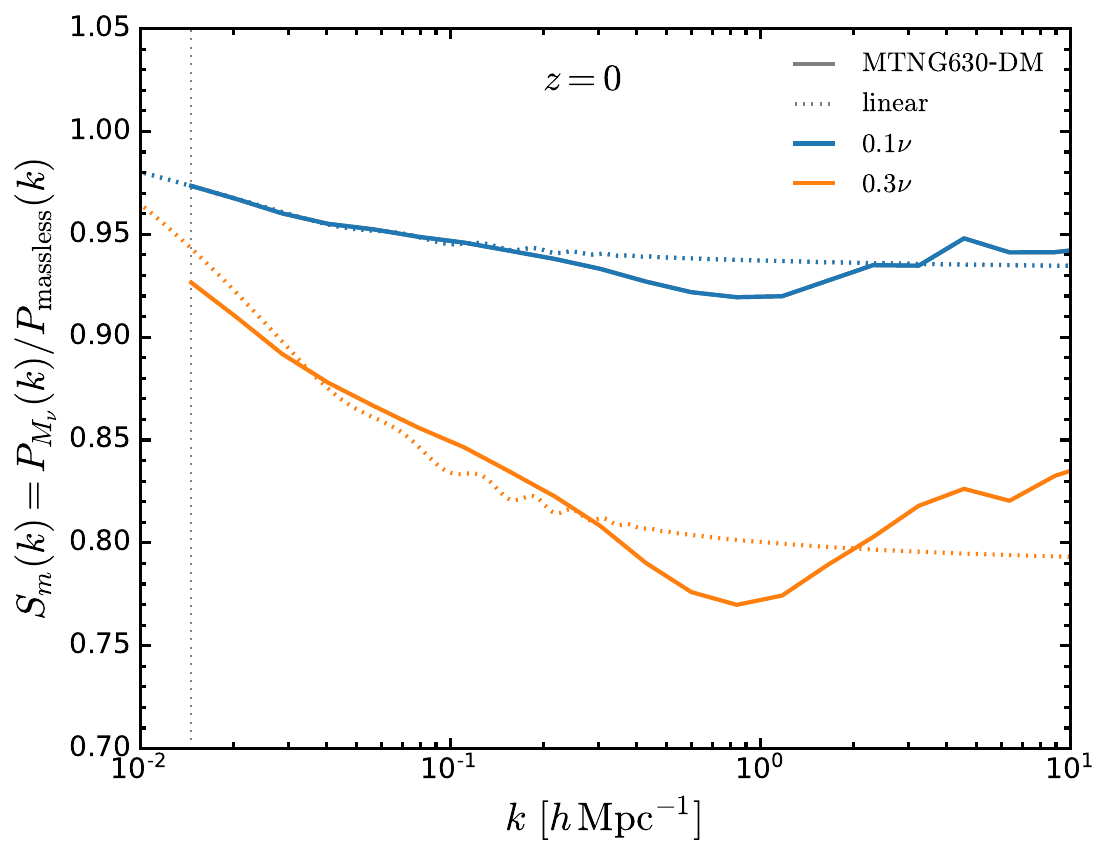}
\includegraphics[width=0.47\textwidth]{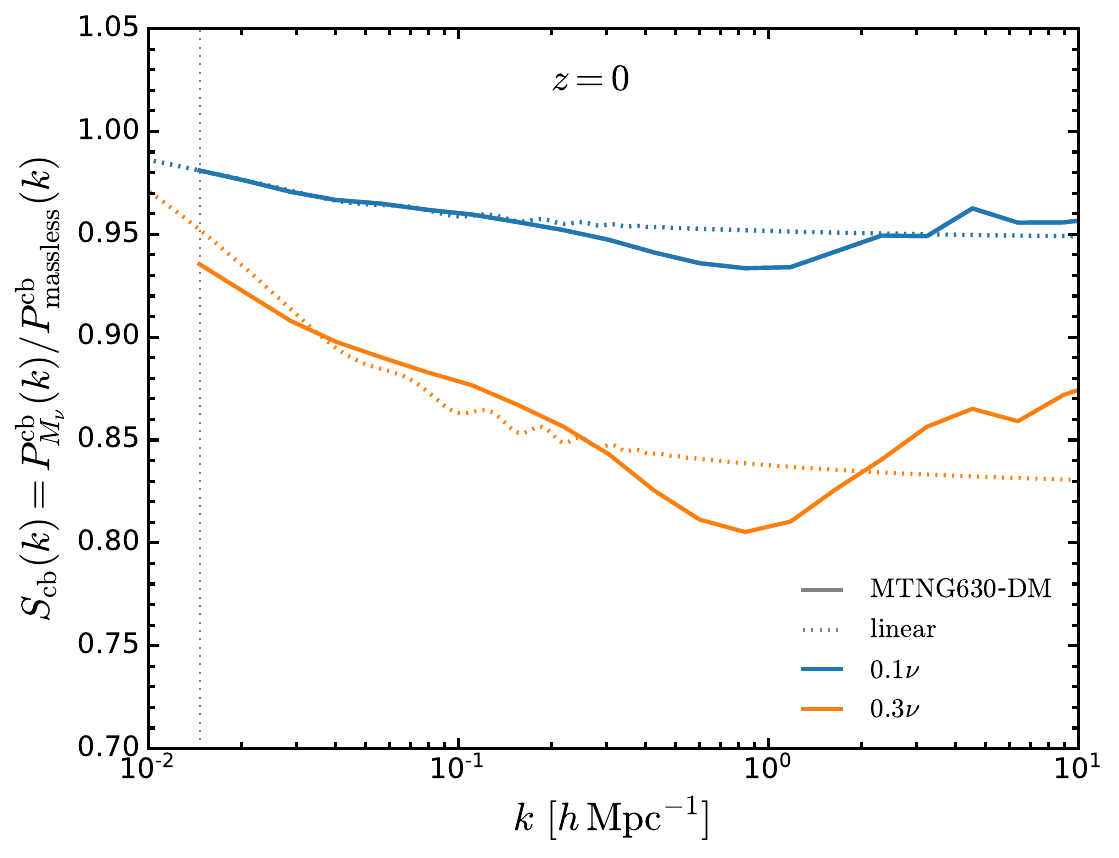}
\caption{Neutrino induced suppression of the total matter (left panel) and cold matter (CDM and baryons; right panels) power spectra at $z=0$ from our MTNG630-DM-$0.1\nu$ (blue curves) and MTNG630-DM-$0.3\nu$ (orange curves) simulations. The dotted curves represent the linear theory prediction from {\small CAMB}. The fundamental mode of the box is indicated by the vertical dotted line.}
  \label{fig:Sknu}
\end{figure*}

The impact of massive neutrinos on the matter clustering is often quantified by comparing the power spectrum with massive neutrinos to the massless ($M_\nu = 0\,{\rm  meV}$) case. This comparison is then expressed in terms of a suppression ratio given by
\begin{equation}\label{eq:Sx}
    S_{\rm X}(k) = \frac{P^{\rm X}_{M_\nu}(k)}{P^{\rm X}_{\rm massless}(k)}\,,
\end{equation}
where X is here a placeholder for the component of interest, and $M_\nu$ corresponds to the total mass of the  neutrino species, i.e.~$100\,{\rm meV}$ or $300\,{\rm meV}$ in our case. In the following we explore this suppression factor for the cold (cb) and total (m) matter power spectra. 

In Figure~\ref{fig:Sknu} we show the suppression induced by massive neutrinos ($M_\nu= 100\,{\rm  meV}$ and $M_\nu= 300\,{\rm  meV}$) in the total matter ($S_{\rm m}(k)$, left panel) and cold matter ($S_{\rm cb}(k)$, right panel) components with respect to the massless case at $z=0$.  We also include the linear theory suppression for each model, for comparison (dotted lines).

On large scales, the suppression predicted by the simulations is in good agreement with the linear theory for both contributions, $P_{\rm m}(k)$ and $P_{\rm cb}(k)$. In the non-linear regime $(0.2<k/[\hMpc]<2)$, however, the suppression takes a ``spoon-like'' shape with a minimum at $\sim 8\%$ for the $M_\nu= 100\,{\rm meV}$ case and $\sim 24 \%$ for the $M_\nu= 300\,{\rm meV}$ case, respectively, for the total matter power spectrum. On the other hand, the cold matter suppression is slightly weaker compared with that for the total matter. For the former we find a $\sim 7\%$ and $\sim 20\%$ suppression at $k\sim 0.8\hMpc$ for the 100 meV and 300 meV models, respectively.

At very small scales, $k > 2\hMpc$, we see an enhancement of power due to effects of nonlinear neutrino clustering. Nevertheless, the impact of massive neutrinos on small-scale structures is weak, meaning that they also are unlikely to significantly affect the abundance of low-mass dark matter haloes (cf. Sec.~\ref{sec:halo}).

\begin{figure}
 \centering
\includegraphics[width=0.48\textwidth]{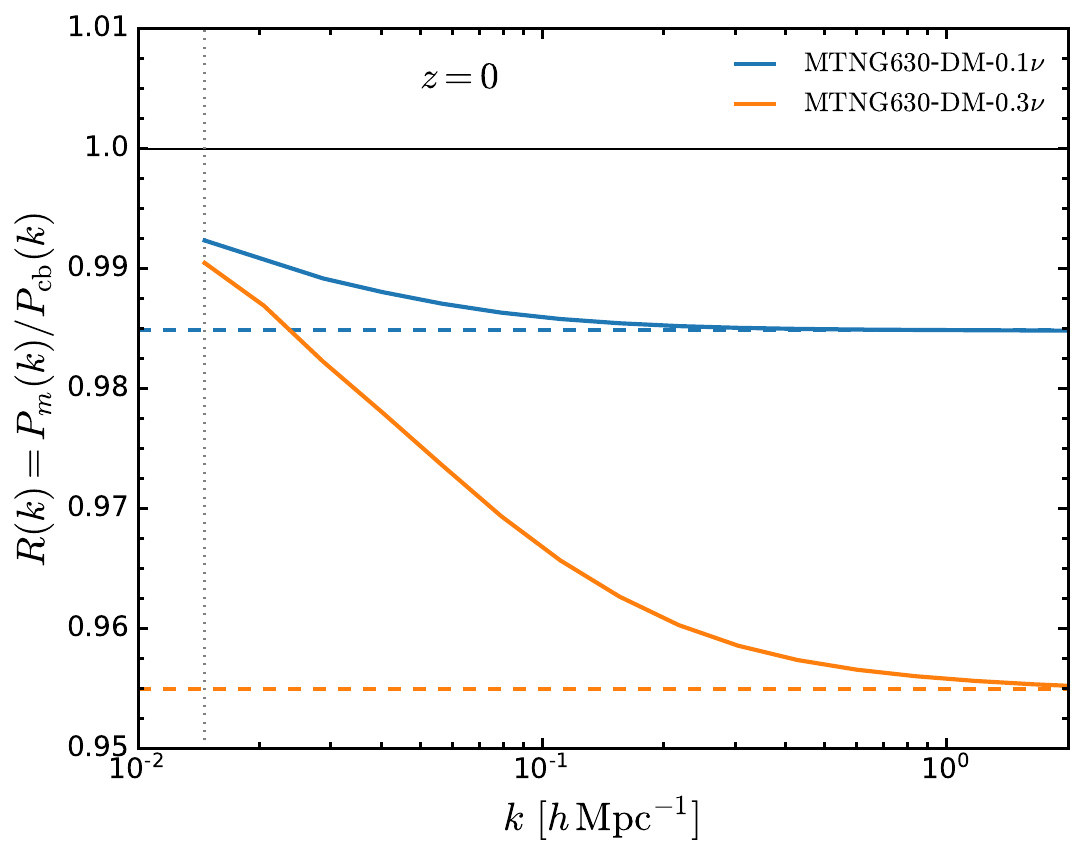}
\caption{Ratio between the total matter and the cold matter power spectra from the MTNG630-DM-0.1$\nu$ (blue solid line) and the MTNG630-DM-0.3$\nu$ (orange solid line) simulations at $z=0$. The horizontal coloured dashed lines indicate the value of $(1-f_\nu)^2$ for each model, and the vertical dotted line represents the fundamental mode of the periodic simulation box.}
  \label{fig:Rk_L430}
\end{figure}

Note that the massless case does not contain a contribution from massive neutrinos, hence $P_{\rm cb}(k) = P_{\rm m}(k)$, and the impact of massive neutrinos on the suppression, Eq.~\eqref{eq:Sx} (see Fig.~\ref{fig:Sknu}), is only driven by $P^{\rm X}_{M_\nu}(k)$. For this reason, we plot in Figure~\ref{fig:Rk_L430}  the ratio between the total matter and the cold matter power spectrum measured from our $100\,{\rm meV}$ and $300\,{\rm meV}$ neutrino simulations. It is expected that this ratio, $R(k) = {P_{\rm m}(k)}/{P_{\rm cb}(k)}$, asymptotes to the values of
\begin{equation}\label{eq:Rknu}
R(k) = \frac{P_{\rm m}(k)}{P_{\rm cb}(k)} \simeq
    \begin{cases}
    1       & \quad \text{for } k \ll k_{\rm nr}\\
    (1-f_\nu)^2  & \quad \text{for } k \gg k_{\rm nr}\,,
  \end{cases}
\end{equation}
where $f_\nu$ is the massive neutrino fraction, defined as $f_\nu \equiv \Omega_\nu / \Omega_{\rm m}$, and $k_{\rm nr}$ is given by \citep{Lesgourgues:2006nd},
\begin{equation}\label{eq:knr}
k_{\rm nr} \simeq 0.018\,\Omega_{\rm m}^{1/2}\left(\frac{1\,{\rm eV}}{M_\nu}\right)\hMpc\,.     
\end{equation}
Our simulation measurements for $R(k)$ are shown in Figure~\ref{fig:Rk_L430} as solid blue and solid orange lines for our MTNG630-DM-0.1$\nu$ and MTNG630-DM-0.3$\nu$ runs at $z=0$, respectively. We can see that the simulations approach the value of unity at sufficiently large-scales $(k \ll k_{\rm nr})$, and they agree perfectly with the asymptotic value of $(1-f_\nu)^2$ on small-scales $(k \gg k_{\rm nr})$. This also explains the $\sim 1.5\%$ and $\sim 4.5\%$ differences when comparing the power spectrum suppression of the total matter and cold matter of the $100\,{\rm meV}$ and $300\,{\rm meV}$ neutrino models shown in Fig.~\ref{fig:Sknu}, respectively. 

\section{Halo statistics}
\label{sec:halo}
\begin{figure}
 \centering
\includegraphics[width=0.48\textwidth]{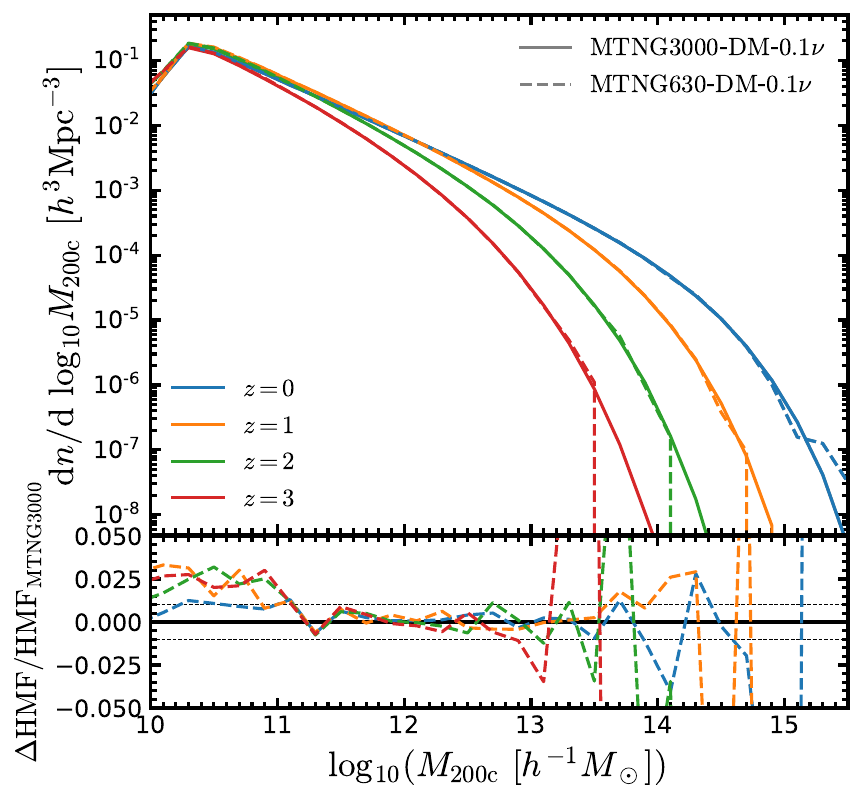}
\caption{Box size impact on the measured differential halo mass function from the MTNG3000-DM-$0.1\nu$ (solid lines) and MTNG630-DM-$0.1\nu$ (dashed lines) simulations at redshifts $z=0$ (blue lines), $z=1$ (orange lines), $z=2$ (green lines) and $z=3$ (red lines). The lower subpanel shows the relative difference with respect to the larger run (MTNG3000), and the horizontal dotted lines in this panel indicate a nominal $1\%$ difference.}
  \label{fig:HMF_100ev}
\end{figure}

\begin{figure*}
 \centering
\includegraphics[width=0.47\textwidth]{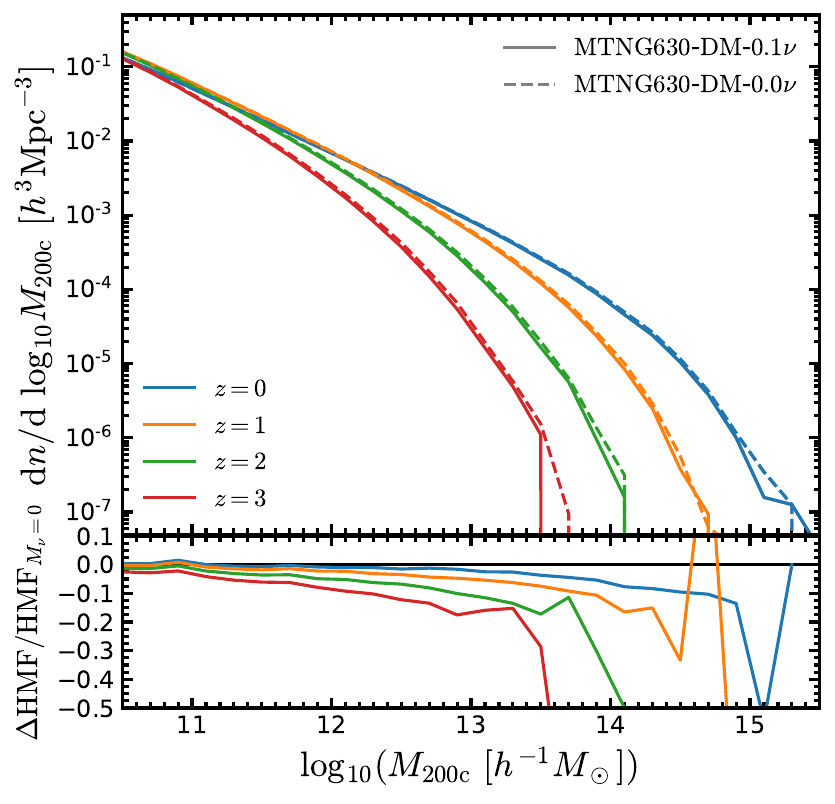}
\includegraphics[width=0.47\textwidth]{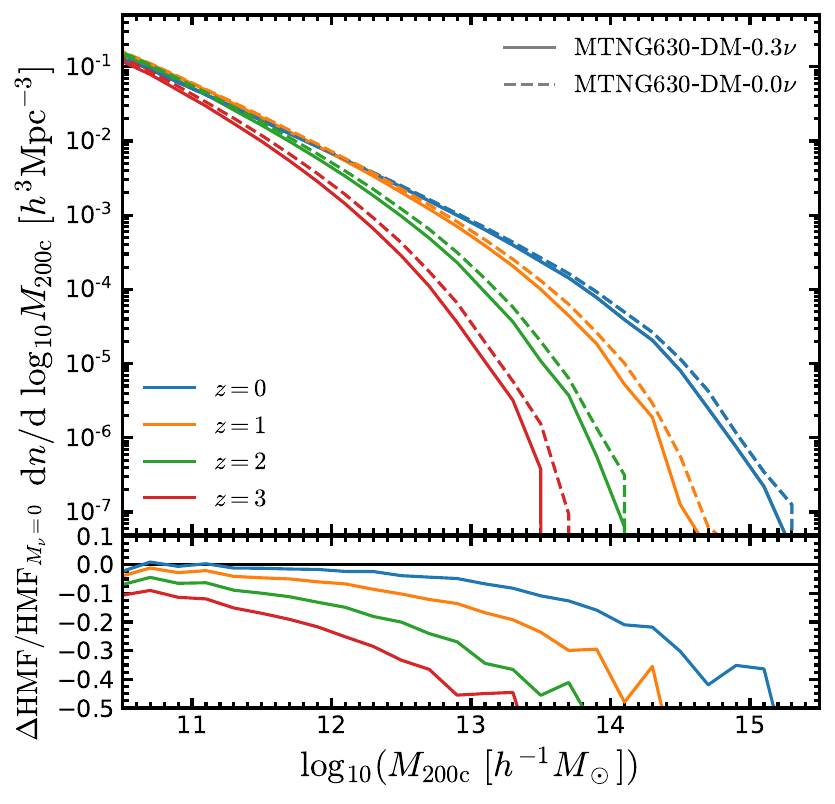}
\caption{Neutrino impact on the halo mass function (HMF) for the neutrino models with $M_\nu = 100\,{\rm meV}$ (left panel) and $M_\nu = 300\,{\rm meV}$ (right panel) at different redshifts, as labelled. In both panels we show the measured HMF for the $\Lambda$CDM model as dashed lines (i.e., for $M_\nu = 0\,{\rm  meV}$). The bottom subpanels display the relative difference between the massive neutrino models and the massless case.}
  \label{fig:HMF_L430}
\end{figure*}

The distribution of cold dark matter (sub)structures is impacted by the presence of massive neutrinos through the scale-dependent suppression of the power spectrum which we have highlighted above. On the other hand, the statistics of cold dark matter haloes, such as galaxy cluster counts, is a powerful probe to constrain cosmology \citep[see e.g.,][]{Costanzi:2013bha,Bohringer:2016fcq}. We thus continue the analysis of the MTNG3000 and MTNG630 simulations by exploring the statistics and clustering of (cold) dark matter haloes in massive neutrino cosmologies. We measure the abundance of haloes from our group catalogues by considering each FoF group with at least one bound subhalo as a main halo, taking the particle with the minimum gravitational potential of the largest subhalo as halo centre. We characterize the halo masses using the $M_{200c}$ spherical overdensity mass definition, i.e.~these masses correspond to the mass contained within a radius enclosing a mean overdensity of 200 times the critical density.

To begin with, we validate the halo mass function (HMF) from our $100\,{\rm meV}$ neutrino simulations. Figure~\ref{fig:HMF_100ev} displays the redshift evolution of the halo mass functions measured as the  mean of the A and B realisations of the MTNG3000-DM-0.1$\nu$ and MTNG630-DM-0.1$\nu$ simulations, at redshifts $z=3$ (red lines), $2$ (green lines), $1$ (orange lines) and $0$ (blue lines). Due to the high resolution of our simulations, we can find well-resolved haloes with masses as low as $M_{200c} \sim 10^{10.6} \Msh$, which corresponds to haloes with 60 CDM particles. Note that this mass limit is well below the halo mass for hosting the emission-line galaxies and ${\rm H}\alpha$ emitters which are the main targets of ongoing and future galaxy surveys, such as DESI and Euclid \citep{Gonzalez-Perez:2017mvf, Merson:2019vfr, Gao:2023osn, Prada:2023lmw, Rocher:2023zyh, DESI:2024rkp, Euclid:2024few}. 

From the lower subpanels of Fig.~\ref{fig:HMF_100ev}, we observe that the MTNG630 runs predict almost the same HMF as the MTNG3000 runs in the mass range $10^{11} < M_{200c}/[\Msh] < 10^{13}$. The high-mass end of the MTNG630 HMF is however affected at all redshifts by the small box size of these runs. This can be seen as increasing counting fluctuations in the higher mass bins, particularly at higher redshift, corresponding to the masses of groups or clusters of galaxies with $M_{200c} > 10^{13}\Msh$, and in an eventual complete lack of the most massive clusters that are present in MTNG3000. However, this test confirms that the MTNG630 runs can still be very useful for cosmological analysis over a wide halo mass range, despite their limited box size.

Next we investigate the impact of varying the mass of neutrinos on the halo population. We plot the evolution of the HMF from $z=3$ to $z=0$ for our MTNG630-DM-0.1$\nu$ and MTNG630-DM-0.3$\nu$ simulations in the left and right panels of Figure~\ref{fig:HMF_L430}, respectively. We additionally include the evolution of the massless neutrino model (MTNG630-DM-0.0$\nu$; dashed lines) in each panel, for comparison.

As expected, the models with massive neutrinos exhibit a reduction of the HMF at the massive end. This well-known suppression of the halo mass function \citep{Castorina:2015bma,Liu:2017now,Adamek2023} is highlighted in the bottom subpanels of Fig.~\ref{fig:HMF_L430}. We observe that the suppression is stronger for higher neutrino mass and higher redshift. For instance, at low masses ($M_{200c} < 10^{12} \Msh$), the suppression due to neutrinos with masses $M_\nu = 100\,{\rm meV}$ (left panel) is less than 10 percent, while for the more massive case (300 meV; right panel) the suppression exceeds 20 percent. On the other hand, for massive haloes with $M_{200c} > 10^{13}\Msh$ there is a deficit of haloes amounting to around 20\% for the $100\,{\rm meV}$ case and even 50\% for the $300\,{\rm meV}$ model. Naturally, this can be expected to have a substantial impact on the galaxy population as well, as we will explore below.

\begin{figure*}
 \centering
\includegraphics[width=0.47\textwidth]{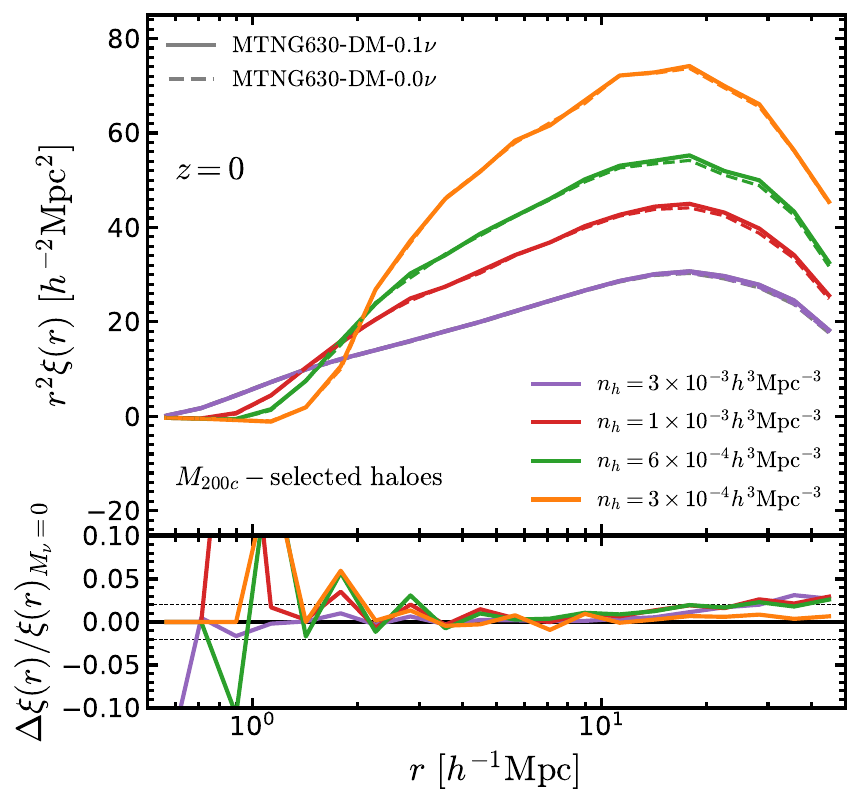}
\includegraphics[width=0.47\textwidth]{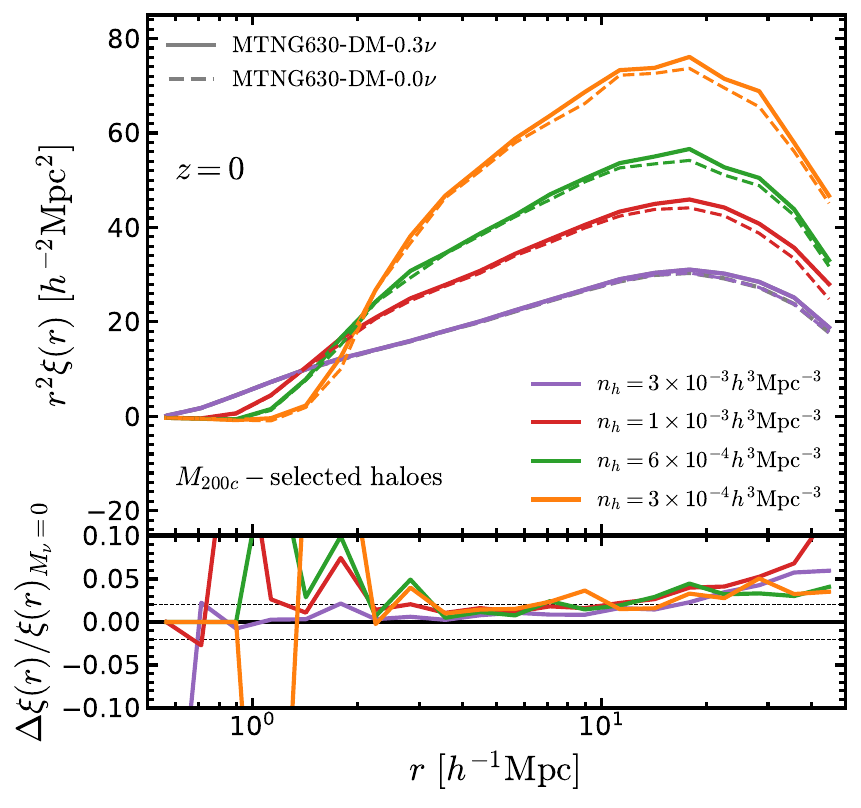}
\caption{Real-space halo two-point correlation function for $M_{200c}$-selected haloes with number densities $n_h = 3\times10^{-3}\Mpchc$ (Nh1, purple lines), $n_h = 1\times10^{-3}\Mpchc$ (Nh2, red lines), $n_h = 6\times10^{-4}\Mpchc$ (Nh3, green lines) and $n_h = 3\times10^{-4}\Mpchc$ (Nh4, orange lines) at $z=0$. The lower subpanels show the neutrino impact on the halo clustering for the $M_\nu = 100\,{\rm meV}$  and $M_\nu = 300\,{\rm meV}$ models, respectively, with the horizontal dashed lines indicating a 2\% difference.}
  \label{fig:xih_L430}
\end{figure*}

Before turning to this, we first investigate the impact of massive neutrinos on the halo clustering of the MTNG630 runs, but for conciseness we only focus on the present time. For definiteness we define four halo populations with number densities of $n_h = 3\times 10^{-3}\Mpchc$ (Nh1 sample), $n_h = 1\times 10^{-3}\Mpchc$ (Nh2 sample), $n_h = 6\times 10^{-4}\Mpchc$ (Nh3 sample) and $n_h = 3\times 10^{-4}\Mpchc$ (Nh4 sample), for the three neutrino models (massless, $100\,{\rm meV}$ and $300\,{\rm meV}$). The specific number densities of our halo selected samples are motivated by the observed number densities of galaxies of recent and current surveys, such as BOSS and DESI \citep{BOSS:2016wmc,DESI:2023bgx}.

Figure~\ref{fig:xih_L430} shows the real-space halo two-point correlation function measured from the MTNG630-DM-0.1$\nu$ (left panel; solid lines), MTNG630-DM-0.3$\nu$ (right panel; solid lines), and MTNG630-DM-0.0$\nu$ (both panels; dashed lines) simulations at $z=0$, for the four halo samples described above. We measured the clustering using 20 log-spaced radial bins between $0.5 < r/[\Mpch] < 50$. Note that we again plot the average of the A and B realisations of each model. We clearly observe from the upper left and right panels that samples dominated by more massive haloes have a larger clustering amplitude at $r > 2\Mpch$, independently of neutrino mass, i.e., samples with lower spatial-density are more clustered and more highly biased at larger separations $r$. This follows the general expectation for an increase of halo bias with halo mass \citep{Mo1996}.

In order to quantify the impact of massive neutrinos on halo clustering we plot the relative difference between the massive neutrino models and the massless case in the lower subpanels of Figure~\ref{fig:xih_L430}. From the $100\,{\rm meV}$ model (left lower panel) we find an enhancement of 3 percent of the clustering on large scales $r>20\Mpch$ for the Nh1, Nh2 and Nh3 halo samples. The clustering of the Nh4 sample agrees very well with its massless counterpart, with a difference of less than 1 percent for scales $r>2\Mpch$. On the other hand, the cold dark matter haloes in the 300 meV massive neutrinos cosmology are more clustered on large-scales $(r>10\Mpch)$, producing an enhancement larger than 2 percent, independently of the halo number density selection. This indicates that the large-scale halo bias in such massive models is higher than in the plain $\Lambda$CDM (massless) model, as previously found by \cite{Villaescusa-Navarro:2013pva} and \cite{Adamek2023}. 

We note that the massive neutrino simulations predict a slightly higher clustering amplitude at small-scales ($r<2\Mpch$) for the Nh2, Nh3 and Nh4 samples as well, for both massive neutrino models.  To match the number of (cold) dark matter haloes of their massless counterpart simulation, each of the corresponding samples needs to include more haloes of lower mass, which are in principle expected to be less clustered, at least on large scales. This slightly counterintuitive result could thus reflect a stronger small-scale clustering of these haloes in neutrino models, or it may also  arise due to changes in the halo bias in massive neutrino cosmologies.

\begin{figure}
 \centering
\includegraphics[width=0.5\textwidth]{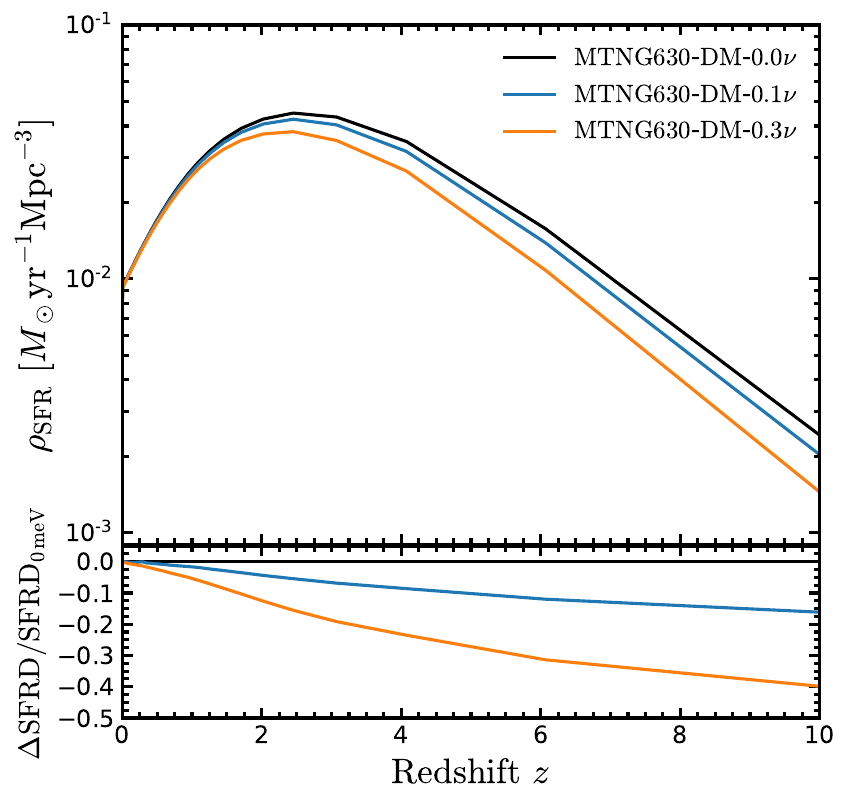}
\caption{Evolution of the cosmic star-formation rate density (SFRD) from our MTNG630 simulations, comparing $0.0\,{\rm meV}$ (black line), $100\,{\rm meV}$ (blue line), and $300\,{\rm meV}$ models (orange line). The lower panel shows the relative difference between the massive neutrinos models and the massless case.}
  \label{fig:SFRD_nu}
\end{figure}

\section{Galaxy statistics}
\label{sec:gal}
\begin{figure*}
 \centering
\includegraphics[width=0.49\textwidth]{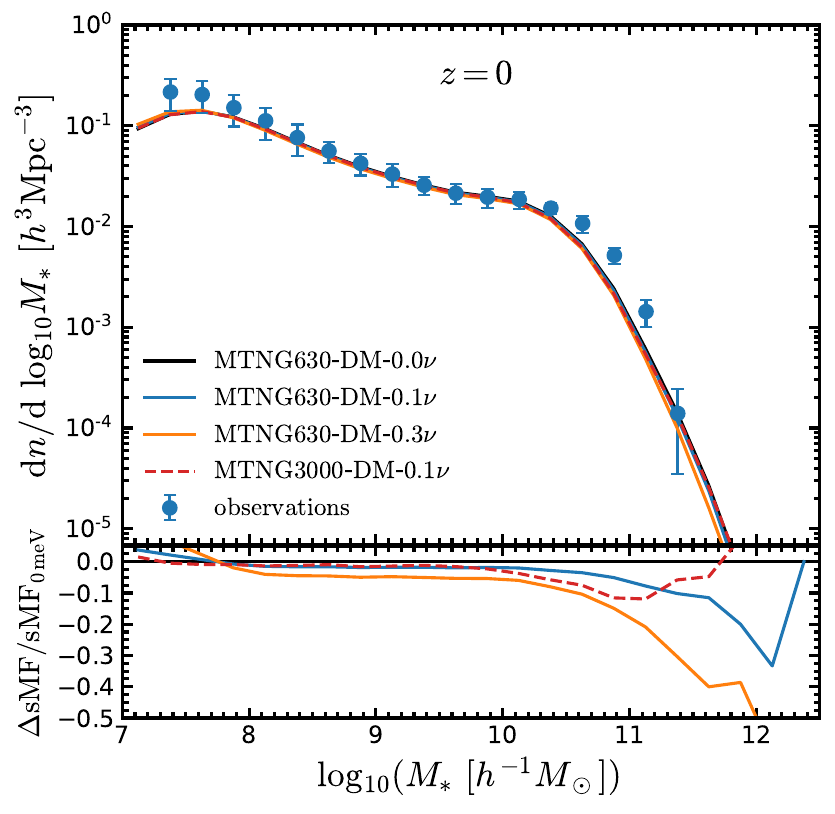}
\includegraphics[width=0.49\textwidth]{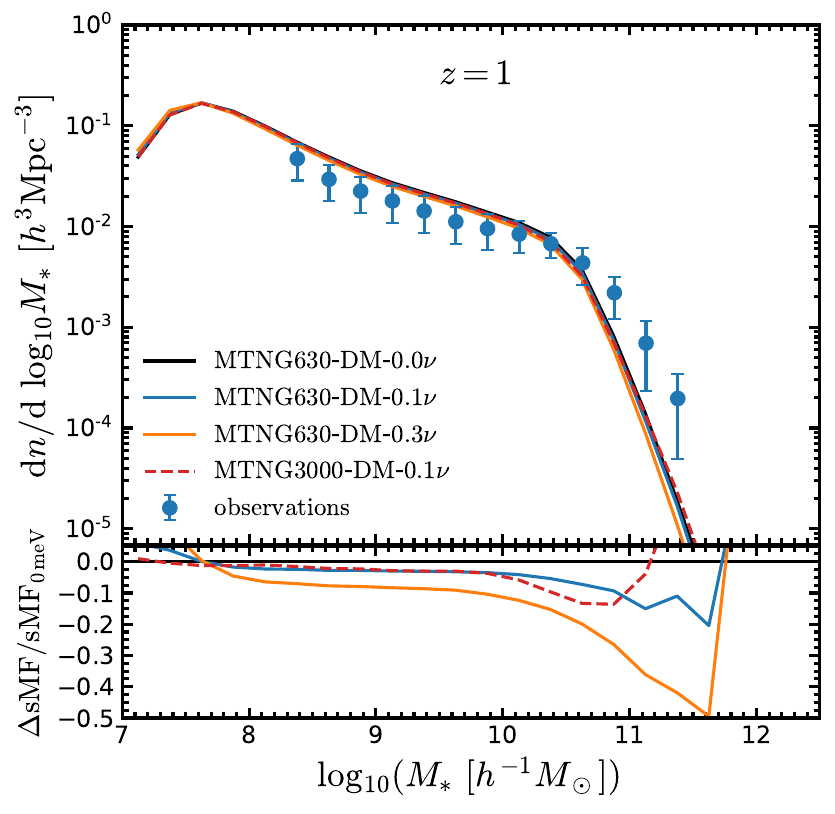}
\caption{Galaxy stellar mass functions at $z=0$ (left panel) and $z=1$ (right panel) from the MTN630 and MTNG3000 neutrino simulations, as labelled. The lower subpanels display the relative difference between the 100 meV and 300 meV models and the massless case. The blue dots with error bars correspond to combined observational data sets presented by \citealt{Henriques:2014sga}.}
  \label{fig:gSMF_nu}
\end{figure*}

In the following, we study the impact of massive neutrinos on the properties and the distribution of galaxies computed with a physics-based semi-analytical galaxy formation model (SAM), something that has not been done so far. We generate realistic galaxy catalogues using the updated version of the {\small L-GALAXIES} semi-analytic galaxy formation code introduced by \citealt{Barrera2023}. This code originated in work done for the Millennium simulation \citep{Springel2005, Croton2006} and has been progressively refined over the years in many studies \citep[e.g.][]{Henriques2013, Henriques2015, Ayromlou2021}. The recent improvements most relevant for the present work include the use of the more robust substructure and merger tree tracking algorithms introduced in {\small GADGET-4}, a reworked time integration scheme of the differential equations used to describe the galaxy formation physics, a more flexible storage scheme for the star formation history of galaxies, and the ability to output galaxies and their photometry directly on the past backwards lightcone of a fiducial observer.

We have run the {\small L-GALAXIES} code on the merger trees of our simulations without any modifications of the galaxy formation physics parameters (such as star formation efficiency, feedback strength. etc.), i.e.~the same values for them in the parameter files and the same compilation flags are used for the three massive neutrino models (0, 100 and 300 meV), modulo the small adjustments in the cosmological parameters of the models needed to realise a massive neutrino component, which has a very small influence on the Hubble rate computation done by {\small L-GALAXIES}. We are thus not testing whether differences due to the presence of neutrinos could be partially or fully absorbed into changes of the (uncertain) galaxy formation physics parameters. Notice that throughout this section we will show results for the average of the A and B realisations, unless stated otherwise.

First, we consider the evolution of the cosmic star formation density (SFRD) for the full galaxy population as measured from the SAM galaxy catalogues of the MTNG630 simulations. In Figure~\ref{fig:SFRD_nu} we show the evolution of the star-formation rate density from $z=10$ to $z=0$. We can clearly see that the massive neutrinos models ($100\,{\rm meV}$ and $300\,{\rm meV}$) track the overall shape of the massless case, i.e., the SFRD first increases rapidly with time (decreasing redshift), reaches a peak at $z\sim2.5$, and then decreases at later times to the present epoch as the dark matter halo population starts to become increasingly dominated by massive quenched galaxies with older stellar populations. However, the SFRD at high redshift is clearly reduced for the massive neutrino models compared to the massless case, whereas the models reach very similar star formation densities at the present time.

We directly quantify the relative impact of massive neutrinos on the SFRD in the lower panel of Fig.~\ref{fig:SFRD_nu}. The star-formation rate of the $M_\nu = 0.3\,{\rm eV}$ model is suppressed at almost all redshifts, but this suppression is much stronger at early times ($z>5$), reaching forty percent with respect to the massless case. This is due to the suppression of low-mass haloes at high redshift in this model (as shown in the right panel of Fig.~\ref{fig:HMF_L430}; there is already a deficit of 30\%  in the CDM haloes with masses $M_{200c} \sim 10^{12}\Msh$). While this effect is weaker for the $100\,{\rm meV}$ neutrino model, it still suppresses the star-formation density by 15\% at redshifts $z>5$. The suppression becomes smaller for both neutrino models at late times, and at low redshift the halo population which accounts for the majority of the star formation is almost the same as in the massless neutrinos counterpart (see, e.g., the orange and blue lines in the lower subpanels of Fig.~\ref{fig:HMF_L430}).

\begin{table*}
\begin{tabular}{ccccccc}
\hline
            & $M_\star$ $[\Msh]$   & $M_\star$ $[\Msh]$   & $M_\star$ $[\Msh]$   & SFR $[\Msyr]$ & SFR $[\Msyr]$ & SFR $[\Msyr]$ \\
            & 0 meV                & 100 meV              & 300 meV              & 0 meV         & 100 meV       & 300 meV       \\ \hline
$z=0$ (Ng1) & $2.82\times 10^{10}$ & $2.74\times10^{10}$  & $2.60\times 10^{10}$ & $2.44$        & $2.45$        & $2.46$        \\
$z=0$ (Ng2) & $5.11\times 10^{10}$ & $4.97\times 10^{10}$ & $4.71\times 10^{10}$ & $4.02$        & $4.04$        & $4.06$        \\
$z=1$ (Ng1) & $1.83\times 10^{10}$ & $1.75\times 10^{10}$ & $1.58\times 10^{10}$ & $7.01$        & $6.92$        & $6.73$        \\
$z=1$ (Ng2) & $3.40\times 10^{10}$ & $3.28\times 10^{10}$ & $3.04\times 10^{10}$ & $1.22$        & $1.22$        & $1.20$        \\ \hline
\end{tabular}
\caption{Stellar-mass and star-formation rate cuts that define the $n_g = 3\times 10^{-3}\hMpcc$ (Ng1) and $n_g = 1\times 10^{-3}\hMpcc$ (Ng2) galaxy samples.}
\label{tab:gal_cuts}
\end{table*}

\begin{figure*}
 \centering
\includegraphics[width=0.47\textwidth]{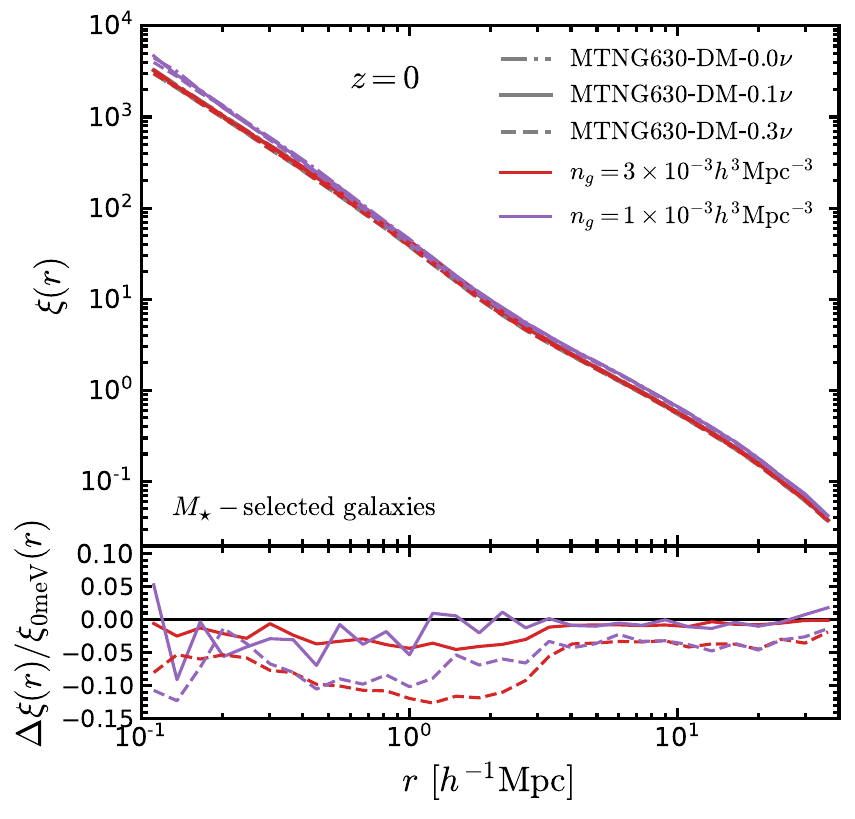}
\includegraphics[width=0.47\textwidth]{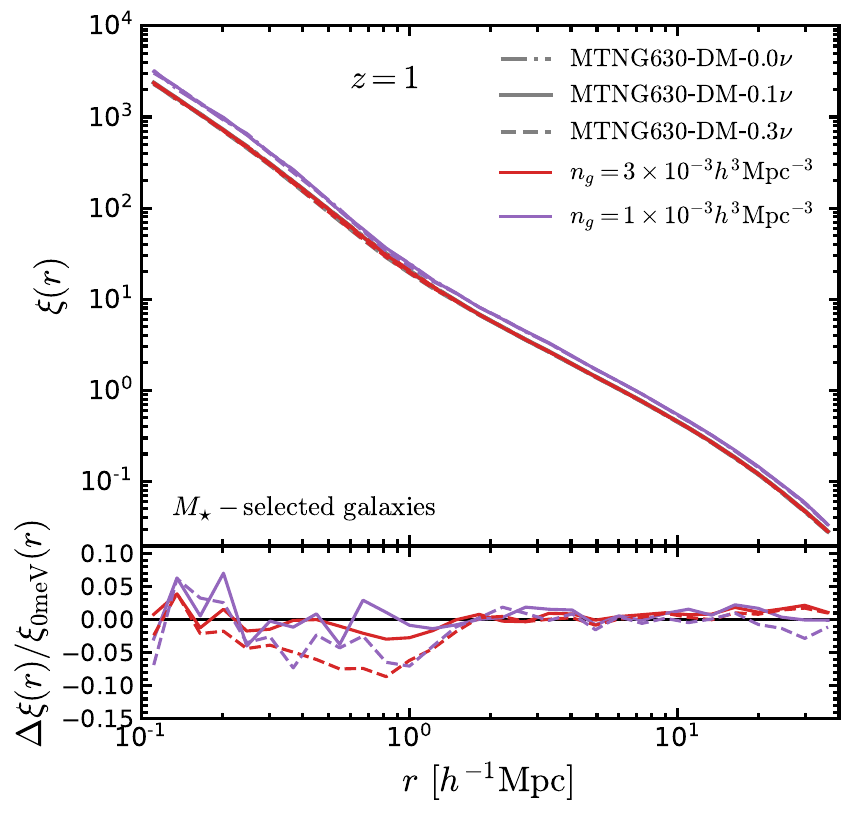}
\includegraphics[width=0.47\textwidth]{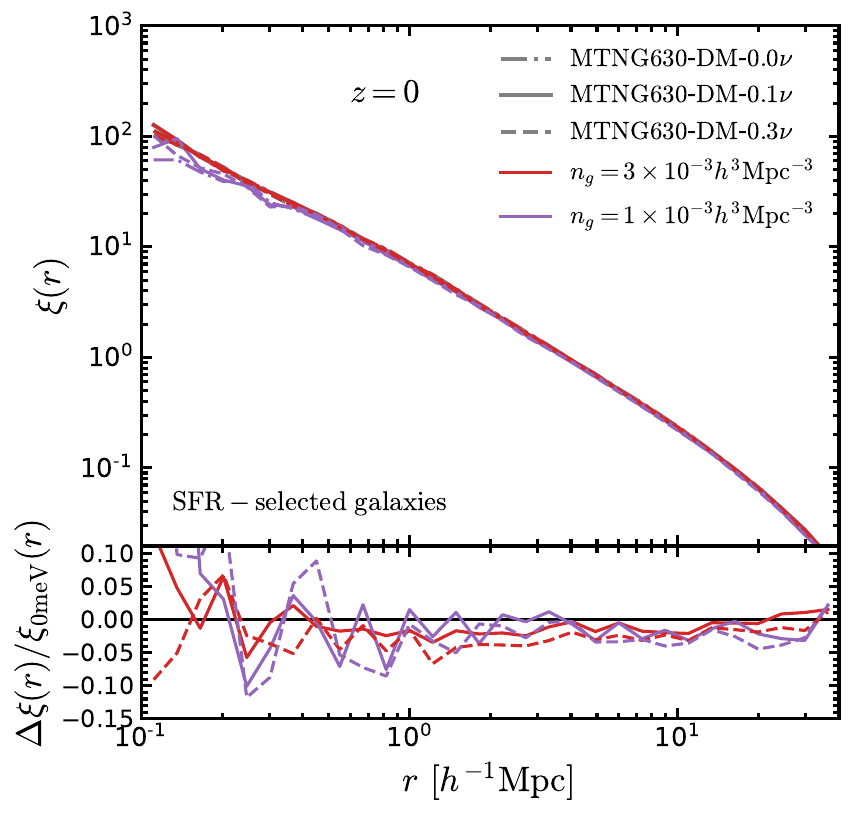}
\includegraphics[width=0.47\textwidth]{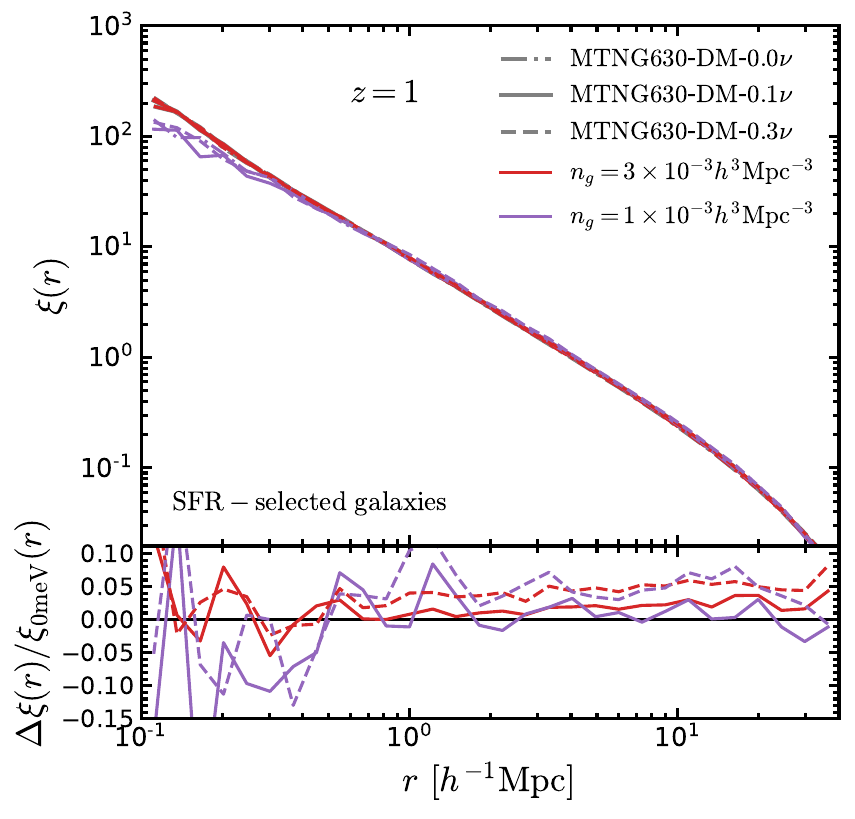}
\caption{Real-space galaxy two-point correlation functions measured from the MTNG630-DM-0.0$\nu$ (dash-dotted lines), MTNG630-DM-0.1$\nu$ (solid lines) and MTNG630-DM-0.3$\nu$ (dashed lines) simulations at $z=0$ (left panels) and $z=1$ (right panels), for stellar mass (upper row) and star-formation rate (bottom row) selected galaxy samples with number densities of $n_g = 3\times 10^{-3}\hMpcc$ (Ng1; red lines) and $n_g = 1\times 10^{-3}\hMpcc$ (Ng2; purple lines). The lower subpanels show the relative differences between the massive (100 meV and 300 meV) and massless (0 meV) models.}
  \label{fig:xigg_L430}
\end{figure*}

We now investigate the galaxy stellar mass function, which is central for describing the formation and evolution of the galaxy population over cosmic time. The galaxy stellar mass function (GSMF) measured from our MTNG630 (solid lines) and MTNG3000 (red dashed lines) simulations is shown in Figure~\ref{fig:gSMF_nu} at $z=0$ (left panel) and $z=1$ (right panel). The GSMFs are generally in good agreement with observations (blue symbols with errorbars) at both redshifts. Also, we see that the resolution of our simulations is sufficient to produce galaxies with a stellar mass of $M_\star = 10^{8}\Msh$, consistent with the convergence studies of \citet{Barrera2023}.

From the bottom subpanels of Fig.~\ref{fig:gSMF_nu}, which display the relative differences between the models, we observe that the stellar mass function measured from the MTNG630-DM-$0.1\nu$ simulation (blue solid line) is in excellent agreement with the massless case for galaxies with $M_\star < 10^{10}\Msh$. However, we find a suppression of the GSMF for galaxies with stellar mass $M_\star \sim 10^{11}\Msh$ of around 10\%, and this suppression is stronger for massive galaxies, reaching 30\% for galaxies with $M_\star = 10^{12}\Msh$ at $z=0$, whereas at the higher redshift of  $z=1$ it is about 20\% for galaxies with $M_\star \sim 10^{11.5}\Msh$. 

Note that we have also included measurements of the GSMF for the MTNG3000 runs (red dashed lines) in all panels of Figure~\ref{fig:gSMF_nu}. For the stellar mass range $10^8 < M_\star/[\Msh] < 10^{10}$ , the suppression of the GSMF predicted by the MTNG3000-DM-0.1$\nu$ simulations is in very good agreement with that inferred from their 630 Mpc counterparts (MTNG630-DM-0.1$\nu$; blue lines). However, for massive galaxies beyond the knee of the GSMF the models for the 3000 Mpc box size predict a higher galaxy abundance. But note that we are in this figure comparing the MTNG3000-DM-0.1$\nu$ ($M_\nu = 0.1$ eV model)  GSMFs runs to the MTNG630-DM-$0.0\nu$ (massless) model, hence the difference at the high-mass end $(M_\star > 10^{11}\Msh)$ is expected due to the different box sizes. In this regime, the MTNG3000 boxes produce a large number of  massive objects that are simply absent in the MTNG630 boxes, and in this sense the latter are not sufficiently representative for the universe as a whole.

We also find that the $M_\nu = 0.3\,{\rm eV}$ model (orange lines in the lower subpanels of Fig.~\ref{fig:gSMF_nu}) produces a suppression of 10\% and 5\% in the mass range $10^8 < M_\star/[\Msh] < 10^{10}$ at $z=1$ and $z=0$, respectively. There is a noticeable difference of around 40\% at the bright end of the galaxy stellar mass function in both panels of Fig.~\ref{fig:gSMF_nu}, between the $M_\nu = 0.3,{\rm eV}$ and the $M_\nu = 0\,{\rm eV}$ models. This is consistent with our findings for the halo mass function shown in the right panel of Fig.~\ref{fig:HMF_L430}.

The clustering of galaxies is another powerful tool to constrain cosmological models and galaxy formation physics. For this reason we here present a preliminary analysis of the galaxy clustering from our neutrino simulations. This is primarily meant as an illustration of what is possible with our models, whereas a comprehensive analysis of the galaxy clustering properties of the MTNG630 and MTNG3000 simulations is beyond the scope of this paper and  is left for future work.

\begin{figure*}
 \centering
\includegraphics[width=0.47\textwidth]{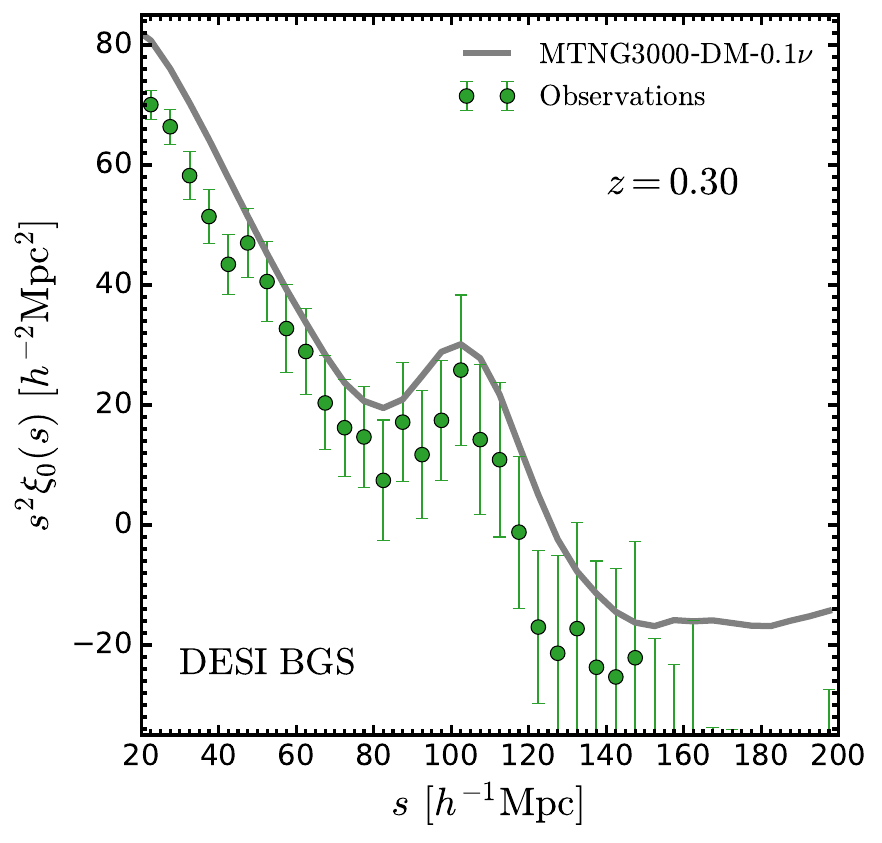}
\includegraphics[width=0.47\textwidth]{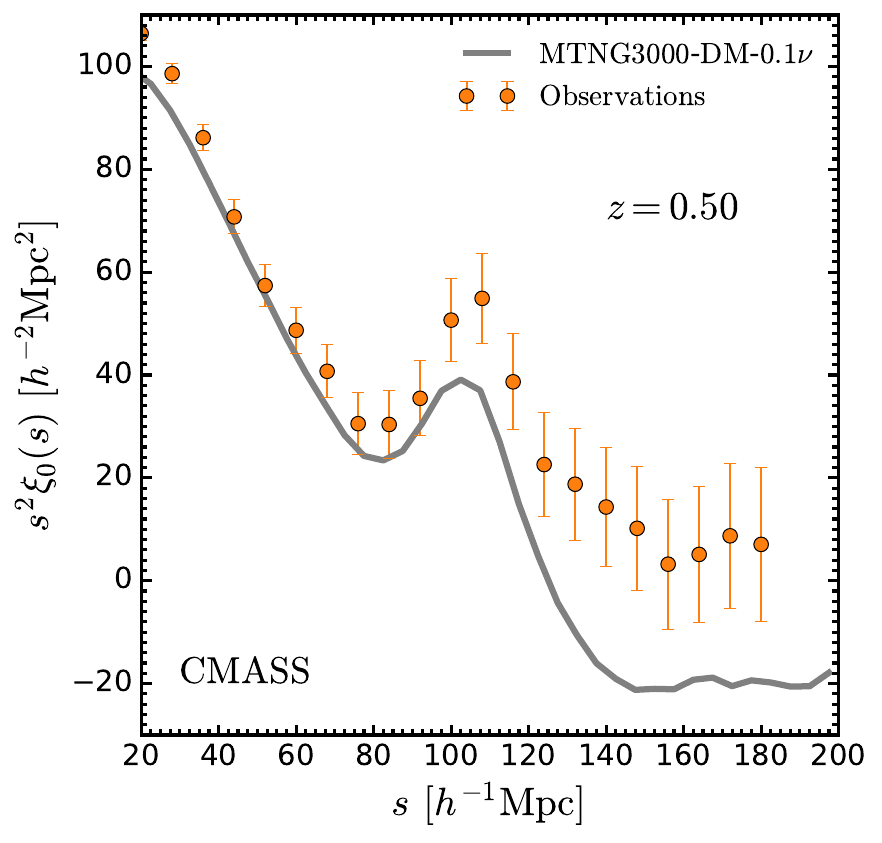}
\includegraphics[width=0.47\textwidth]{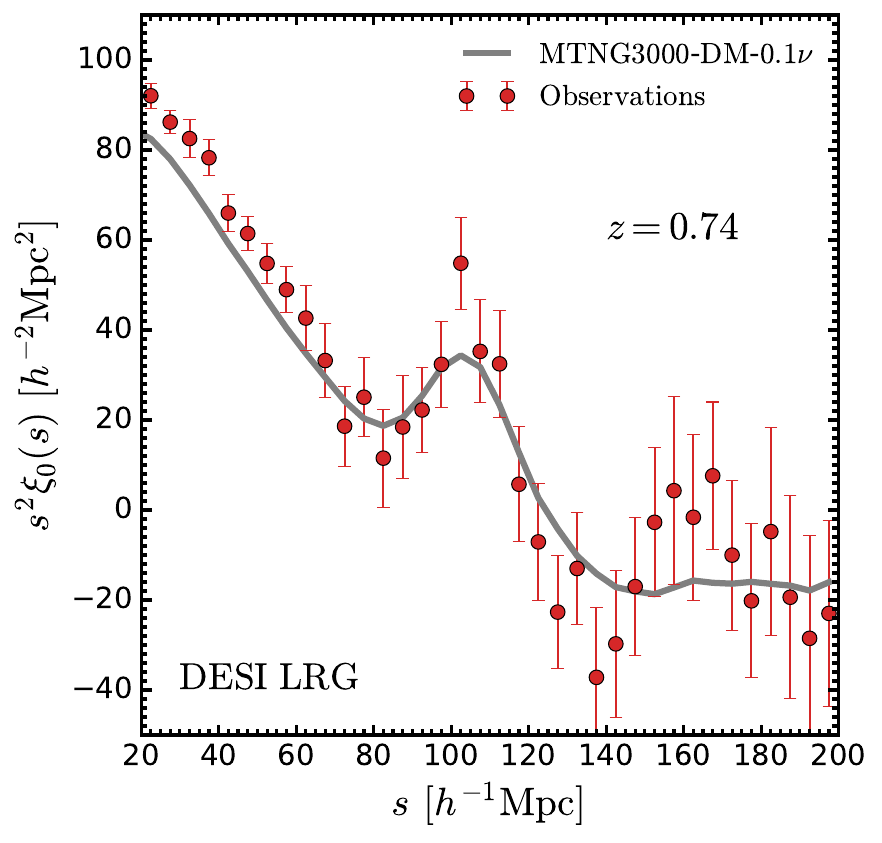}
\includegraphics[width=0.47\textwidth]{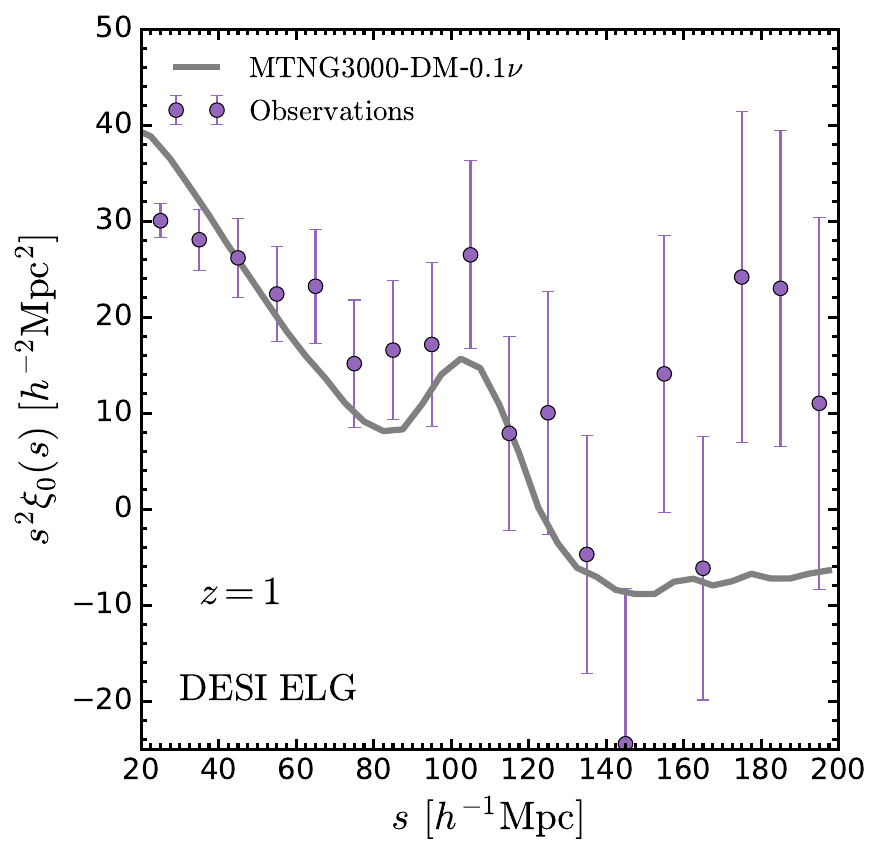}
\caption{Monopole of the redshift-space correlation function measured from the MTNG3000-DM-0.1$\nu$ simulations for different galaxy selections and redshifts, as labelled. Coloured symbols with errorbars are observations from the DESI BGS (green symbols), CMASS (orange symbols), DESI LRG (red symbols) and DESI ELG (purple symbols) surveys. The CMASS observational data points were taken from \citealt{BOSS:2015dhs} and the BGS, LRG and ELG data correspond to the DESI-M2 sample by \citealt{DESI:2023bgx}.}
  \label{fig:xi_L2040}
\end{figure*}

We measure the real-space galaxy two-point correlation function (2PCF) from the MTNG630 simulations at $z=1$ and $z=0$ using 30 logarithmically spaced radial bins in the range $0.1 < r/[\Mpch] < 40$, as presented in Fig.~\ref{fig:xigg_L430}. We consider both stellar mass $(M_\star)$ and star-formation rate (SFR) selected galaxies with number densities of $n_g = 3\times 10^{-3}\hMpcc$ (Ng1) and $n_g=1\times 10^{-3}\hMpcc$ (Ng2). The selection by $M_\star$ and SFR was chosen because it has been shown that these quantities are good proxies to define luminous-red and emission-line galaxies in the context of galaxy surveys \citep[see e.g.,][]{Hadzhiyska:2020iri,Bose2023,Hadzhiyska2023a}. The number densities of the Ng1 and Ng2 samples are similar to the expected spatial densities of galaxies that will be observed by the DESI and Euclid surveys. For definiteness, the specific values of the stellar mass and star formation cuts that we adopted to yield these number densities are listed in Table~\ref{tab:gal_cuts}.

The galaxy 2PCF measured from the $M_\star$-selected samples is shown in the upper row of Figure~\ref{fig:xigg_L430}. From the upper right panel of Fig.~\ref{fig:xigg_L430} we see that the three neutrino models predict almost the same clustering at $z=1$ at scales of $r>2\Mpch$ for both galaxy samples,  Ng1 and Ng2. On small-scales, $r<1\Mpch$, the massive neutrinos suppress the clustering signal by around 2\% and 7\%, for the $100\,{\rm meV}$ and $300\,{\rm meV}$ models, respectively. On the other hand, at $z=0$ (upper left panel), the clustering signal of galaxies from the $100\,{\rm meV}$ $M_\star-$samples is similar to the results at $z=1$, because the galaxy stellar mass function is only slightly affected by the 100 meV neutrinos at $z=0$ and $z=1$ (cf. Figure~\ref{fig:gSMF_nu}). In the 300 meV model, we find a deviation of 5\% below the massless case at scales $r>3\Mpch$ at $z=0$. Larger effects are seen on scales $r<2\Mpch$, where the clustering signal drops by around 15\% and 10\% compared to the massless model for the Ng1 and Ng2 samples, respectively. 

The clustering of SFR-selected galaxies is displayed in the bottom row of Fig.~\ref{fig:xigg_L430}. We find that the clustering signal of SFR-selected galaxies is generally weaker than for the stellar-mass selected samples. This is because star-forming galaxies are often blue, young, low-mass galaxies that are less clustered than massive galaxies. We observe that the 2PCF of the $100\,{\rm  meV}$ model at $z=1$ (bottom right subpanel of Fig.~\ref{fig:xigg_L430}) is 2\% larger than that of the massless case at scales $r>1\Mpch$. The amplitude of the galaxy clustering signal of the $300\,{\rm meV}$ samples at $z=1$ is higher than for the massless case. This can be understood because we need to include more galaxies with lower SFR to match the desired number density.  At the present time (bottom left subpanel of Fig.~\ref{fig:xigg_L430}), the clustering amplitude predicted by the massive neutrino models is $3\%$ lower than for the massless case. This is because at the present time the Ng1 and Ng2 samples have almost the same SFR cut (see Table~\ref{tab:gal_cuts}) for all models. Note that in both panels we observe that the SFR-Ng2 sample of the $100\,{\rm meV}$ model (solid purple line) produces almost the same clustering amplitude as the massless case over a broad range of scales. This is also expected based on the nearly identical SFR cut in both models. Nevertheless, there is a noticeable difference on small-scales, $r<0.3\Mpch$, where the amplitude is decreased by 10\%.

Finally, we can take advantage of the large volume of the MTNG3000-DM-0.1$\nu$ simulations to measure the large-scale baryonic acoustic oscillation (BAO) scale galaxy clustering from our SAM catalogues. One of the goals of ongoing and future galaxy surveys is to use the BAO feature as seen in different galaxy selections across cosmic time to reconstruct the cosmic expansion history, and in this way to constrain the cosmological parameters -- ideally including the sum of the neutrino masses.

As an illustration of the expected effects in this approach, we measure the monopole of the redshift-space correlation function at different redshifts using our MTNG3000 runs and compare the results with current observational data. To do so, we define four galaxy samples that match the observed galaxy number densities, as follows:
\begin{itemize}
    \item {\bf DESI BGS}: the desired number density is $n_{\rm BGS} = 3.8\times 10^{-3}\hMpcc$ at the median redshift of $z=0.3$ \citep{DESI:2023bgx}. We select galaxies ranked by their $r-$band magnitude from our {\small L-GALAXIES} catalogues, matching the observed $n_{\rm BGS}$.
    \item {\bf CMASS}: this sample has an observed spatial number density of $n_{\rm CMASS} = 4.4\times 10^{-4}\hMpcc$ at $z=0.5$ \citep{BOSS:2015dhs}. In this case, we rank galaxies in our simulations by stellar mass in descending order, and we select the $N$ most massive galaxies that reproduce the observational number density.
    \item {\bf DESI LRG}: similarly to the CMASS sample, we rank the SAM galaxies by stellar mass, and then select them to reproduce a number density of $n_{\rm LRG} = 7.4\times 10^{-4}\hMpcc$ at $z=0.74$ \citep{DESI:2023bgx}.
    \item {\bf DESI ELG}: to build this sample we just apply a simple SFR-cut in our galaxy catalogues such that we match the observed number density of $n_{\rm ELG} = 1\times 10^{-3}\hMpcc$ at $z=1$ \citep{DESI:2023bgx}.
\end{itemize}

The resulting clustering measurements from our simulations are shown in Figure~\ref{fig:xi_L2040}. We observe that our simple galaxy selections based on $r-$band magnitude, galaxy stellar mass, and star-formation rate reproduce the amplitude and shape of the observed galaxy clustering signal quite well, which is reassuring. A potentially sizeable mismatch is seen for CMASS at scales $ s > 120\,h^{-1}{\rm Mpc}$, but whether or not this is really significant can only be assessed with a more careful comparison based on a detailed forward modelling approach.

To build the fully realistic mock galaxy catalogues required for this exercise, we need to apply a more sophisticated colour-magnitude selection and impose survey masks to also account for the appropriate survey geometry \citep[see e.g.,][]{BOSS:2016wmc, Raichoor:2022jab, DESI:2022gle} and the effects this entails. However, as our goal here is merely to show that our MTNG3000 simulations have sufficiently large volume to produce galaxy catalogues that are capable of such BAO-scale analysis we defer an in-depth analysis of the match to observational galaxy surveys to a dedicated future study that implements a detailed forward modelling.

\section{Summary and conclusions}
\label{sec:conc}
We have introduced a new set of simulations that are part of the MillenniumTNG (MTNG) project. These simulations focus on the neutrino sector and consist in each case of two realisations (A and B) for cosmic variance reduction. We include massive neutrinos with summed rest masses $M_\nu = 100\,{\rm meV}$ in two different volumes with box sizes $L=2040\Mpch$ (3000 Mpc) and $L=430\Mpch$ ($\sim$630 Mpc), dubbed  MTNG3000-DM-0.1$\nu$ and MTNG630-DM-0.1$\nu$, respectively. The large box runs evolve close to 1.1 trillion cold dark matter particles, and more than 16 billion neutrino particles with the $\delta f$-technique, giving the cold dark matter a particle mass resolution of $6.66\times 10^8 \Msh$. In addition, we have carried out two further simulations for neutrino models with masses $M_\nu = 300\,{\rm meV}$ and $M_\nu = 0\,{\rm meV}$, using the smaller volume configuration (MTNG630-DM-0.3$\nu$ and MTNG630-DM-0.0$\nu$).

In this paper, we have also detailed the numerical implementation of our approaches to include massive neutrinos (and other relativistic species) in cosmological simulations. This entailed changes in the {\small GADGET-4} code to properly account for relativistic components, and the use of the so-called $\delta f$-method \citep{Elbers2021} to reduce shot noise in a particle-based representation of neutrinos, as well as adjustments of the initial conditions generation to be able to directly align them with the multi-species transfer functions computed by Boltzmann codes such as {\small CAMB}, and to correct for super-horizon modes present in $\sim$Gpc-scale N-body simulations.

As part of a series of validation tests of the matter clustering of our $M_\nu = 100\, {\rm meV}$ simulations, we have found an excellent agreement between the cold matter (CDM+b) and massive neutrinos components of the measured power spectrum from MTNG3000-DM-0.1$\nu$ and  MTNG630-DM-0.1$\nu$ (see Figure~\ref{fig:Pk_box}). In addition, we established that the evolution of the matter clustering of the MTNG630-DM-0.1$\nu$ and MTNG630-DM-0.3$\nu$ simulations is in good agreement with linear theory predictions on the largest scales (see Fig.~\ref{fig:Pkm_L430}). Also, we observed a {\it spoon}-like suppression in the power spectrum due to presence of massive neutrinos (in Fig.~\ref{fig:Sknu}). The corresponding effect is stronger for more massive neutrino models, e.g., $M_\nu = 300\, {\rm meV}$, and the small-scale behaviour of the effect reflects non-linear neutrino clustering.

We have then explored the statistics and clustering signal of cold dark matter haloes in our simulations with massive neutrinos. We find excellent agreement between the halo mass functions measured from the MTNG3000-DM-0.1$\nu$ and MTNG630-DM-0.1$\nu$ simulations (see Fig.~\ref{fig:HMF_100ev}). We also quantified the impact of massive neutrinos on the halo mass function (HMF), finding a deficit of haloes that becomes larger for more massive neutrino models, and is stronger at higher redshifts (see Fig.~\ref{fig:HMF_L430}). Massive neutrinos also affect the clustering of cold dark matter haloes, where we detect noticeable differences at large scales $r>10-20\Mpch$, depending on the neutrino model (cf. Fig.~\ref{fig:xih_L430}).

Finally, one of the most important advances in the MTNG simulation project is that it can produce physically grounded galaxy catalogues with the updated version of the {\small L-GALAXIES} semi-analytical model of galaxy formation \citep{Barrera2023}. With such galaxy catalogues we are able to study the impact of massive neutrinos on astrophysical and cosmological observables. For instance, we showed that massive neutrinos suppress the star-formation history of galaxies across cosmic time (cf. Fig.~\ref{fig:SFRD_nu}). In addition, the galaxy stellar mass function (GSMF; Fig.~\ref{fig:gSMF_nu}) is affected  in a similar fashion to the HMF, i.e., the predicted GSMF in models with massive neutrinos is lower than that of their massless counterpart. Moreover, we established some deviations in the real-space galaxy correlation functions of stellar mass and star formation rate selected galaxies in different neutrino cosmologies (see Fig.~\ref{fig:xigg_L430}). Furthermore, using our large boxes (MTNG3000-A/B) we were able to measure the clustering of galaxies at the BAO scale. We find that simple galaxy selection criteria that yield similar galaxy densities as in the observed samples can reproduce the shape of the large-scale galaxy clustering of current observation rather well and reproduce a clear BAO feature with very high signal to noise (as shown in Fig.~\ref{fig:xi_L2040}).

The transformative increase of the observational volume and the number of observed galaxies ushered in by galaxy surveys such as DESI and Euclid calls for precise numerical calculations to assist the proper interpretation of this data. The MTNG simulation project is an attempt to live up to this challenge and to provide a useful basis for the generation of truly realistic galaxy mock catalogues that can improve the faithfulness of comparisons between theory and observations. While the present paper has merely scratched the surface in this regard, it established important foundations for this goal by demonstrating that simulations with massive neutrinos can be carried out with high accuracy at low computational cost, and that their effects should thus be included in future studies of precision cosmology with simulations. While the corresponding effects due to neutrinos are admittedly small, they are nevertheless large enough to offer the exciting prospect to put tighter constraints on the neutrino masses based on cosmological observations -- which is addressing a truly fundamental goal of astroparticle physics.

\section*{Acknowledgements}
CH-A acknowledges support from the Excellence Cluster ORIGINS which is funded by the Deutsche Forschungsgemeinschaft (DFG, German Research Foundation) under Germany's Excellence Strategy -- EXC-2094 -- 390783311. 
VS and LH acknowledge support by the Simons Collaboration on ``Learning the Universe''. 
LH is supported by NSF grant AST-1815978.  
SB is supported by the UK Research and Innovation (UKRI) Future Leaders Fellowship [grant number MR/V023381/1]. 
RK acknowledges support of the Natural Sciences and Engineering Research Council of Canada (NSERC) through a Discovery Grant and a Discovery Launch Supplement, funding reference numbers RGPIN-2024-06222 and DGECR-2024-00144.
The authors gratefully acknowledge the Gauss Centre for Supercomputing (GCS) for providing computing time on the GCS Supercomputer SuperMUC-NG at the Leibniz Supercomputing Centre (LRZ) in Garching, Germany, under project pn34mo. 
This work used the DiRAC@Durham facility managed by the Institute for Computational Cosmology on behalf of the STFC DiRAC HPC Facility, with equipment funded by BEIS capital funding via STFC capital grants ST/K00042X/1, ST/P002293/1, ST/R002371/1 and ST/S002502/1, Durham University and STFC operations grant ST/R000832/1. 

\section*{Data Availability}
The MillenniumTNG simulations will be made fully publicly available at \url{https://www.mtng-project.org} in 2024. The data underlying this article will be shared upon reasonable request to the corresponding authors.




\bibliographystyle{mnras}
\bibliography{ref} 




\bsp	
\label{lastpage}
\end{document}